\documentclass{JHEP3}
\usepackage{amsmath,amssymb,amsfonts}
\usepackage{epsfig}
\usepackage{lscape}
\usepackage{multirow}
\usepackage{cite}

\makeatletter

\@addtoreset{equation}{section}

\newcommand\Ref[1]     {Ref.\,\cite{#1}}
\newcommand\Refs[1]    {Refs.\,\cite{#1}}
\newcommand\eqn[1]     {Eq.\,(\ref{#1})}
\newcommand\eqns[2]    {Eqs.\,(\ref{#1}) and~(\ref{#2})}
\newcommand\eqnss[2]   {Eqs.\,(\ref{#1})--(\ref{#2})}
\newcommand\fig[1]     {Fig.\,{\ref{#1}}}
\newcommand\sect[1]    {Sect.\,{\ref{#1}}}
\newcommand\sects[2]   {Sects.\,\ref{#1} and~\ref{#2}}
\newcommand\sectss[2]  {Sects.\,\ref{#1}--\ref{#2}}

\newcommand\tab[1]     {Tab.~\ref{#1}}
\newcommand\nn         {\nonumber}

\def\beq{\begin{equation}}
\def\eeq{\end{equation}}
\def\bsp#1\esp{\begin{split}#1\end{split}}
\def\beeq{\begin{eqnarray}}
\def\eeeq{\end{eqnarray}}

\renewcommand\i        {{\mathrm i}}
\renewcommand\O        {{\mathrm O}}
\newcommand\Oe[1]      {\ensuremath{\mathrm O(\ep^{#1})}}
\newcommand{\ep}       {\epsilon}
\newcommand\Li         {\mathop{\mathrm{Li}}\nolimits}

\newcommand{\rd}       {{\mathrm{d}}}
\newcommand{\PS}[1]    {\rd\phi_{#1}}

\newcommand{\cI}       {{\cal I}}
\newcommand{\cJ}       {{\cal J}}
\newcommand{\cK}       {{\cal K}}

\newcommand{\cE}       {{\cal E}}
\newcommand{\cF}       {{\cal F}}
\newcommand{\cG}       {{\cal G}}
\newcommand{\cH}       {{\cal H}}
\newcommand{\cII}[1] {{\cal I}\kern-3pt *\kern-3pt{\cal I}_{#1}}
\newcommand{\cIJ}     {{\cal I}\kern-3pt *\kern-3pt{\cal J}}
\newcommand{\cJJ}[1] {{\cal J}\kern-3pt *\kern-3pt{\cal J}_{#1}}
\newcommand{\cJI}     {{\cal J}\kern-3pt *\kern-3pt{\cal I}}
\newcommand{\cKJ}[1] {{\cal K}\kern-3pt *\kern-3pt{\cal J}_{#1}}
\newcommand{\cKI}     {{\cal K}\kern-3pt *\kern-3pt{\cal I}}

\newcommand{\ti}[1]    {\tilde{#1}}

\newcommand{\mom}[1]   {\{p\}^{#1}}
\newcommand{\momt}[1]   {\{\ti{p}\}^{#1}}
\newcommand{\Y}[2]     {Y_{\ti{#1}\ti{#2},Q}}
\newcommand{\cmap}[1]   {\stackrel{{\rm C}_{#1}}{\longrightarrow}}
\newcommand{\smap}[1]   {\stackrel{{\rm S}_{#1}}{\longrightarrow}}

\title{
Analytic integration of real-virtual counterterms
in NNLO jet cross sections II}

\author{Paolo Bolzoni\\
DESY \\
Platanenalle 6, D-15738 Zeuthen, Germany\\
E-mail: \email{paolo.bolzoni@desy.de}}

\author{Sven-Olaf Moch \\
DESY\\
Platanenalle 6, D-15738 Zeuthen, Germany\\
E-mail: \email{sven-olaf.moch@desy.de}}

\author{G\'abor Somogyi \\
Institute for Theoretical Physics, University of
Z\"urich\\ Winterthurerstrasse 190, CH-8057 Z\"urich, Switzerland\\
E-mail: \email{sgabi@physik.unizh.ch}}

\author{Zolt\'an Tr\'ocs\'anyi\\
University of Debrecen and Institute of Nuclear Research of the 
Hungarian Academy of Sciences, H-4001 Debrecen P.O.Box 51, Hungary\\
E-mail: \email{z.trocsanyi@atomki.hu}}

\abstract{
We present analytic expressions of all integrals required to 
complete the explicit evaluation of the real-virtual integrated
counterterms needed to define a recently proposed subtraction scheme
for jet cross sections at next-to-next-to-leading order in QCD. 
We use the Mellin-Barnes representation of these integrals in $4-2\ep$
dimensions to obtain the coefficients of their Laurent expansions around $\ep=0$. 
These coefficients are given by linear combinations of multidimensional
Mellin-Barnes integrals.  
We compute the coefficients of such expansions in $\ep$ 
both numerically and analytically by complex integration over the Mellin-Barnes contours.
}

\preprint{arXiv:0905.4390 [hep-ph] \\
DESY 09-075 \\
SFB/CPP-09-42 \\
ZU-TH 07/09}

\begin{document}


\section{Introduction}
\label{sec:intro}

Precision predictions in perturbative Quantum Chromodynamics (QCD) at colliders demand
calculating physical observables beyond leading order (LO) accuracy and, 
in the traditional approach to higher order predictions with 
fully differential kinematics, real and virtual corrections are separately evaluated. 
Integration over the phase space then requires a consistent treatment of the
infrared singularities before any numerical computation may be performed.  
At next-to-leading order (NLO), infrared divergences can be handled 
using a subtraction scheme, which exploits the universal 
structure of the kinematical singularities of QCD matrix elements.
The necessary (process-independent) counterterms regularize
the virtual corrections at one loop and the real emission phase space integrals
simultaneously~\cite{Catani:1996vz}.

At next-to-next-to-leading order (NNLO), the calculation of the radiative
corrections to fully differential cross sections is a challenging problem 
and various extensions of the subtraction method at NNLO have been proposed, 
see e.g. \Refs{%
GehrmannDeRidder:2004tv,Weinzierl:2003fx,Frixione:2004is,Somogyi:2005xz}. 
Currently, the available results for electron-positron annihilation at NNLO 
include total rates~\cite{GehrmannDe Ridder:2007bj,GehrmannDeRidder:2008ug,Weinzierl:2008iv}
and event shapes~\cite{GehrmannDeRidder:2007hr,Weinzierl:2009ms}
for the process $e^+ e^- \to 3~$jets and are all based on the antenna subtraction
method~\cite{GehrmannDeRidder:2005aw,GehrmannDeRidder:2005hi,GehrmannDeRidder:2005cm}.
On the other hand for colorless final states, such as vector boson or Higgs boson production at hadron colliders 
dedicated subtraction schemes at NNLO~\cite{Catani:2007vq,Catani:2009sm} have been applied.
The infrared structure of scattering processes with three or more colored partons is involved 
if calculated at NNLO with the antenna subtraction method~\cite{Weinzierl:2009nz}~-- 
a fact which has motivated the formulation of alternative subtraction schemes.
In particular, \Refs{Somogyi:2006cz,Somogyi:2006db,Somogyi:2006da} introduce 
a scheme for computing NNLO corrections to QCD jet cross sections for processes 
without colored partons in the initial state and an arbitrary number 
of massless particles (colored or colorless) in the final state. 
Very recently, following the steps of \Ref{Somogyi:2006cz}, 
this subtraction scheme has been extended to cross sections for hadron-initiated processes~\cite{Somogyi:2009ri}, 
although yet to NLO accuracy only, but in a way which is NNLO-compatible.

Any subtraction scheme is of practical utility only after the counterterms for the regularization of
the real emissions are integrated over the phase space of the unresolved partons. 
In the scheme of \Refs{Somogyi:2006cz,Somogyi:2006db,Somogyi:2006da} these counterterms are universal 
and, therefore can be computed once and for all.
Their knowledge is necessary to regularize the infrared divergences appearing 
in the virtual corrections. 
Some of the integrals needed explicitly in the so-called real-virtual counterterms 
of this scheme have been calculated in \Refs{Aglietti:2008fe,Somogyi:2008fc}. 
In the present paper we complete this task by computing all integrals needed
for the the real-virtual counterterms in the subtraction scheme of 
\Refs{Somogyi:2006cz,Somogyi:2006db,Somogyi:2006da} by means of Mellin-Barnes (MB) representations.
The use of MB integrals when dealing with Feynman integral calculus has proved powerful 
in the last years.
MB integrals were first applied to Feynman integrals in \Refs{Bergere:1973fq,Usyukina:1975yg} 
and pioneering work has been performed since then in \Refs{Boos:1990rg,Smirnov:1999gc,Tausk:1999vh} 
(see also \Ref{Smirnov:2004ym} and references therein for many other examples).
For a given integral the MB representation replaces the power of a sum in the integrand 
by a product of the individual terms of the sum raised to some other powers. 
This leads then to integrals over certain complex contours of $\Gamma$-functions.  
As a crucial point it is then very convenient with this MB representation 
to resolve all singularities in the limit $\ep=0$ within dimensional regularization, $d=4-2\ep$.  
In this paper, we adapt the MB method to derive analytic expressions for 
all integrals appearing in the real-virtual counterterms of \Refs{Somogyi:2006cz,Somogyi:2006db,Somogyi:2006da}.

Let us briefly discuss the merits of the analytic approach for the
computation of the integrated subtraction terms.
First of all, in a higher-order computation, the $\ep$ poles of the 
integrated subtraction terms need to cancel the corresponding $\ep$ poles 
coming from the loop matrix elements in the virtual corrections. 
The cancellation of these poles can be demonstrated 
most convincingly once the pole structure of the integrated subtraction 
terms is exhibited analytically. 
Second, in terms of speed and precision of the evaluation, analytic results 
are very fast and very accurate compared to numerical ones. 
Moreover, they demonstrate that the final result consists of smooth functions only.
Nevertheless also the numerical evaluation of the integrated counterterms has its utility, 
because it serves as an independent check. 
Then, there are indeed some cases, where it is very difficult to find the analytic computation 
of the multi-dimensional MB integral and only the complex numerical integration 
can be carried out. 
In these cases, however, the method of MB integrals provides a fast and reliable way 
to obtain the final results with small numerical uncertainties. 
From a practical point of view, the combination of both, analytic and
numerical evaluations of all MB integrals implies that the final results for
the integrated real-virtual counterterms can be conveniently given e.g.\ in the
form of interpolating tables which can be computed once and for all. 
This suffices for any practical application, because in an actual computation 
the relative uncertainty associated with 
the numerical phase space integrations is generally much greater than that
of the integrated subtraction terms. 

The outline of the paper is the following.
In \sect{sec:Subtractions} we briefly review the phase space integrals of the
real-virtual corrections at NNLO and we define the integrals of the subtraction terms 
that we will consider in this paper.
In \sect{sec:Method} we present a brief explanation of the method of MB representations.
We outline the steps of our calculation and we also 
discuss explicitly an example to display the typical structure of the integrals we are interested in. 
In \sect{sec:Iints} we complete the analytic evaluation of all integrals needed for
integrated collinear counterterms. 
Next, in \sectss{sec:IIints}{sec:KIint} we compute also all different types of the nested integrals.
Finally in \sect{sec:conclusion} we present the conclusions of this work.

\section{Integrals needed for the integrated subtraction terms}
\label{sec:Subtractions}

The subtraction method developed in \Refs{Somogyi:2006da,Somogyi:2006db}
relies on the universal soft and collinear factorization properties of
QCD squared matrix elements. Once the subtraction scheme is defined, one
has to integrate the subtraction terms over the factorized phase space 
of the unresolved parton(s). 
This is the content of the present work (see
also \Ref{Aglietti:2008fe}).

There are two crucial elements in the formulation of a subtraction scheme beyond NLO. 
Firstly, the factorization formulae should disentangle the overlaps in soft-singular 
factors and collinear singularities in order to avoid multiple subtractions
and a simple solution to this problem has been given in \Ref{Nagy:2007mn}.
Secondly, because the factorization formulae are valid only in the strict soft
and collinear limits, they have to be extended to the whole phase space. 
Typically, this requires a mapping of the original $n$ momenta 
$\mom{}_n = \{p_1,\dots, p_n\}$ in an $n$-parton matrix element at any order in perturbation theory 
to $m$ momenta $\momt{}_m = \{\ti p_1,\dots, \ti p_m\}$
in such a way, that momentum conservation is preserved.
Here $m$ denotes the number of hard partons and $n-m$ is the number 
of unresolved  ones.

\FIGURE{
\includegraphics[width=6.75cm]{figs/PSCir.epsi}
\hspace*{10mm}
\includegraphics[width=7.75cm]{figs/PSSr.epsi}
\caption{Graphical representations of the momentum mappings
and the implied phase space factorization: 
collinear (left) and soft momentum mapping (right).
\label{fig:PSCir-PSSr}
}
}

The original $n$-particle phase space of total momentum $Q$ reads
\beq
\PS{n}(p_1,\ldots,p_n;Q) =
\prod_{i=1}^{n}\frac{\rd^d p_i}{(2\pi)^{d-1}}\,\delta_+(p_i^2)\,
(2\pi)^d \delta^{(d)}\left(Q-\sum_{i=1}^{n}p_i\right)
\,,
\label{eq:PSn}
\eeq
and, for a given mapping, one obtains the phase-space factorization as 
\beq
\PS{n}(\mom{}_n;Q)=\PS{m}(\momt{}_{m};Q)
\: [\rd p_{n-m;m}(\mom{}_{n-m};Q)]
\,,
\label{eq:PSfact}
\eeq
which was first introduced in \Ref{Catani:1996vz} in the context of computing
QCD corrections at NLO.
In this paper we are concerned with the integrals of the singly-unresolved counterterms 
(i.e. the case $m=1$), which imply two types of mappings:
\beeq
\mom{}_n&\cmap{ir}&\momt{(ir)}_{n-1} \,=\,
\{\ti{p}_1,\ldots,\ti{p}_{ir},\ldots,\ti{p}_n\}
\,,
\label{eq:cmap} 
\\
\mom{}_n&\smap{r}&\momt{(r)}_{n-1} \,=\, 
\{\ti{p}_1,\ldots,\ti{p}_n\}
\, .
\label{eq:smap}
\eeeq
In the collinear momentum mapping $\cmap{ir}$ in \eqn{eq:cmap} 
the momenta $p_i^\mu$ and $p_r^\mu$ are replaced by a single momentum $\ti{p}_{ir}^\mu$ 
and all other momenta are rescaled, 
while for soft-type subtractions, $\smap{r}$ in \eqn{eq:smap} 
the momentum $p_r^\mu$, that may become soft, is missing from the set, 
and all other momenta are rescaled and transformed by a proper Lorentz transformation. 
Both momentum mappings and the corresponding factorization of the
phase-space measure are represented graphically in \fig{fig:PSCir-PSSr}, 
where the symbol $\otimes$ stands for the convolution as implied by \eqn{eq:PSfact}.
The integration of the singly-unresolved subtraction terms requires three basic types 
of integrals over the corresponding factorized phase space, as well as iterations of these 
(nested integrals are denoted by a $\ast$). 
All necessary integrals were derived in \Refs{Aglietti:2008fe,Somogyi:2008fc}.

\subsection{Basic integrals}
\label{sec:basicintegrals}

The three basic integrals are those used in the collinear, soft and
soft-collinear subtraction counterterms. The collinear integrals have the 
general form
\beeq
&&
\cI(x;\ep,\alpha_0,d_0;\kappa,k,\delta,g_I^{(\pm)}) =
x
\int_0^{\alpha_0}\! \rd \alpha\, \alpha^{-1-(1+\kappa)\ep}
\,(1-\alpha)^{2d_0-1}\,[\alpha+(1-\alpha)x]^{-1-(1+\kappa)\ep}
\nn\\[2mm]&&\qquad
\times
\int_0^1\! \rd v
[v\,(1-v)]^{-\ep}
\left(\frac{\alpha+(1-\alpha)xv}{2\alpha+(1-\alpha)x}\right)^{k+\delta\ep}
\,g_I^{(\pm)}\left(\frac{\alpha+(1-\alpha)xv}{2\alpha+(1-\alpha)x}\right)
\,.
\label{eq:Iint}
\eeeq

\TABLE{
\hspace*{20mm}
\begin{tabular}{|c|c|c|}
\hline
\hline
$\delta$ & Function & $g_I^{(\pm)}(z)$ \\
\hline
  & & \\[-4mm]
$0$ & $g_A$ & $1$ \\[2mm]
$\mp 1$ & $g_B^{(\pm)}$ & $(1-z)^{\pm\ep}$ \\[2mm]
$0$ & $g_C^{(\pm)}$ &
$(1-z)^{\pm\ep}{}_2F_1(\pm\ep,\pm\ep,1\pm\ep,z)$\\[2mm]
$\pm 1$ & $g_D^{(\pm)}$ & ${}_2F_1(\pm\ep,\pm\ep,1\pm\ep,1-z)$ \\[2mm]
\hline
\hline
\end{tabular}
\hspace*{20mm}
\label{tab:I1ints}
\caption{The values of $\delta$ and $g_I^{(\pm)}(z_r)$ for which
\eqn{eq:Iint} needs to be evaluated.}
}

These integrals need to be known as a function of $x \in [0,1]$ (see \Refs{Somogyi:2006cz,Somogyi:2006db,Somogyi:2006da}
for its kinematic definition) in a Laurent-expansion in $\ep$ for $k=-1, 0, 1, 2$.
The necessary values of $\delta$ and the expressions for the functions $g_I^{(\pm)}$ are given in \tab{tab:I1ints}. 
Here $\kappa=0,1$ for the first row and $\kappa=1$ for all other cases. 
Analytic expressions for all cases corresponding to the first two rows of \tab{tab:I1ints} were derived in \Ref{Aglietti:2008fe} 
and contain the first five terms in the $\ep$-expansion.
We compute all cases anew and present our results explicitly in \sect{sec:Iints}. 
The other parameters $\alpha_0 \in (0,1]$ and $d_0$ in \eqn{eq:Iint} will be specified in \sect{sec:Method}.
Our analytic results for these integrals include all the coefficients of the poles in $\ep$ and 
the first three terms in the $\ep$-expansion.

%
%

Next, the soft subtractions require the integral
\beq
\bsp
\cJ(\Y{i}{k};\ep,y_0,d'_0;\kappa) &=
-(4Y_{\ti{i}\ti{k},Q})^{1+\kappa\ep}
\frac{\Gamma^2(1-\ep)}{2\pi\Gamma(1-2\ep)}
\Omega^{(1+\kappa\ep,1+\kappa\ep)}(\cos\chi)\\
&\times
\int_0^{y_0}\!\rd y\,
y^{-1-2(1+\kappa)\ep}(1-y)^{d'_0+\kappa\ep}\,,
\esp
\label{eq:Jint}
\eeq
as a function of $\Y{i}{k} \in [0,1]$ (see \Refs{Somogyi:2006cz,Somogyi:2006db,Somogyi:2006da}
for its kinematic definition) in a Laurent expansion around $\ep = 0$, 
where $\Omega^{(i,k)}(\cos\chi)$ denotes the angular integral in $d$-dimensions
\beq
\bsp
\Omega^{(i,k)}(\cos\chi) &=
\int_{-1}^1\!\rd(\cos\vartheta)\;(\sin\vartheta)^{-2\ep}
\int_{-1}^1\!\rd(\cos\varphi)\;(\sin\varphi)^{-1-2\ep}\\
&\times
(1-\cos\vartheta)^{-i}
(1-\cos\chi\cos\vartheta-\sin\chi\sin\vartheta\cos\varphi)^{-k}
\,,
\esp
\label{eq:Omegajl}
\eeq
with 
\beq
\cos\chi = 1-2\,\Y{i}{k}
\,.
\label{eq:coschi}
\eeq
Finally, the soft-collinear subtractions lead to the integral
\beq
\bsp
\cK(\ep,y_0,d'_0;\kappa) &=
\,2\int_0^{y_0}\! \rd y\,y^{-(2+\kappa)\ep}
(1-y)^{d'_0-1}\int_{-1}^1\! \rd (\cos\vartheta)\,(\sin\vartheta)^{-2\ep}
\label{eq:Kint}
\\ &\times
\left[1+\frac{2(1-y)}{y(1-\cos\vartheta)}\right]^{1+\kappa\ep}
\frac{\Gamma^2(1-\ep)}{2\pi\Gamma(1-2\ep)}
\int_{-1}^{1}\! \rd (\cos\varphi)\,(\sin\varphi)^{-1-2\ep}
\,,
\esp
\eeq
which does not depend on kinematical variables. 
The integrals $\cJ$ and $\cK$ in \eqns{eq:Jint}{eq:Kint} 
have been computed in \Ref{Aglietti:2008fe} for
all relevant parameters in $y_0$, $d'_0$ and $\kappa$. 
We have evaluated these soft and soft-collinear integrals, too, 
and we have checked that the two results agree numerically. 
We do not deal with the cases $\cJ$ and $\cK$ in this paper.

%
%

\subsection{Nested integrals}
\label{sec:iterated}

In a NNLO computation, also iterations of the above integrals appear.
In this paper we complete the list of nested integrals necessary for the 
integrated real-virtual counterterms, in particular we cover all cases 
which have not been addressed in \Ref{Aglietti:2008fe}. 

Of the nested integrals, which we generally denote by a star $*$, 
three are collinear integrals with one of the basic types in its argument,
\beeq
&&
\cII{i}(x;\ep,\alpha_{0},d_{0};k,l) =
x\int_{0}^{\alpha_{0}}
\!\rd\alpha\int_{0}^{1}\!\rd v\,
\alpha^{-1-\ep}\,(1-\alpha)^{2d_{0}-1}\,[\alpha+(1-\alpha)x]^{-1-\ep}
\nn\\[1mm] &&\quad\times
[v(1-v)]^{-\ep}
\left[\frac{\alpha+(1-\alpha)xv}{2\alpha+(1-\alpha)x}\right]^{k}
\cI\Bigg(x\frac{\alpha+(1-\alpha)x(1-v)}{2\alpha+(1-\alpha)x};
\ep,\alpha_0,d_0;0,l,0,1\Bigg)
\,,\qquad~
\label{eq:IIiint}
\eeeq
\beeq
&&
\cII{r}(x;\ep,\alpha_{0},d_{0};k,l) =
x\int_{0}^{\alpha_{0}}
\!\rd\alpha\int_{0}^{1}\!\rd v\,
\alpha^{-1-\ep}\,(1-\alpha)^{2d_{0}-1}\,[\alpha+(1-\alpha)x]^{-1-\ep}
\nn\\[1mm] &&\quad\times
[v(1-v)]^{-\ep}
\left[\frac{\alpha+(1-\alpha)xv}{2\alpha+(1-\alpha)x}\right]^{k}
\cI\Bigg(x\frac{\alpha+(1-\alpha)xv}{2\alpha+(1-\alpha)x};
\ep,\alpha_0,d_0;0,l,0,1\Bigg)
\,,
\label{eq:IIrint}
\eeeq
which we need for $k,\, l = -1, 0, 1, 2$, and
\beeq
\cIJ(x;\ep,\alpha_{0},d_{0},y_{0},d'_{0};k) &&=
x\int_{0}^{\alpha_{0}}
\!\rd\alpha\,\int_{0}^{1}\!\rd v\,
\alpha^{-1-\ep}\,(1-\alpha)^{2d_{0}-1}\,[\alpha+(1-\alpha)x]^{-1-\ep}
\nn\\[1mm] &&\times
[v(1-v)]^{-\ep}
\left[\frac{\alpha+(1-\alpha)xv}{2\alpha+(1-\alpha)x}\right]^{k}
\label{eq:IJint}
\\[1mm] \nn&&\times
\cJ\left(\frac{\alpha(\alpha+(1-\alpha)x)(2\alpha+(1-\alpha)x)^{2}}
{(\alpha+(1-\alpha)xv)(\alpha+(1-\alpha)x(1-v)) \, x^2};
\ep,y_0,d_0',0\right)
\,,
\eeeq
for $k = -1, 0, 1, 2$. 
Both, $\cII{}$ and $\cIJ{}$ are needed as a function of $x \in [0,1]$ in an $\ep$-expansion 
with $\cI$ and $\cJ$ given in \eqns{eq:Iint}{eq:Jint}, respectively. 
A discussion about the choice  of the relevant parameters 
$\alpha_{0}$, $d_{0}$, $y_{0}$ and $d'_{0}$ is given at the end of \sect{sec:Method}  and details of the computation 
are also given in \sect{sec:IIints}.

Three other iterated integrals are defined as soft integrals with other
soft integrals appearing in the argument,
\beq
\bsp
&
\cJJ{ik}(\Y{i}{k};\ep,y_{0},d'_{0}) =
- 8\Y{i}{k}
\,\frac{\Gamma^{2}(1-\ep)}{2\pi\Gamma(1-2\ep)}
\,\int_{-1}^{1}\rd(\cos\vartheta)\,(\sin\vartheta)^{-2\ep}
\\[2mm]
&\qquad\times
\int_{-1}^{1}\rd(\cos\varphi)
\,(\sin\varphi)^{-1-2\ep}(1-\cos\vartheta)^{-1}
\\[1mm]
&\qquad\times
\int_{0}^{y_{0}}\rd y\,y^{-1-2\ep}(1-y)^{d'_{0}}
[2-(1+\cos\chi)\cos\vartheta-\sin\chi\sin\vartheta\cos\varphi]^{-1}
\\[1mm]
&\qquad\qquad\times
\cJ\bigg(\frac{4(1-y)\Y{i}{k}}{[2-y(1+\cos\vartheta)]
[2-y(1+\cos\chi\cos\vartheta+\sin\chi\sin\vartheta\cos\varphi)]};
\ep,y_{0},d'_{0},0
\bigg)
\,,
\esp
\label{eq:JJikint}
\eeq
\beq
\bsp
\cJJ{ir}(\Y{i}{k};\ep,y_{0},d'_{0}) &=
- 8\Y{i}{k}
\,\frac{\Gamma^{2}(1-\ep)}{2\pi\Gamma(1-2\ep)}
\int_{-1}^{1}\rd(\cos\vartheta)\,(\sin\vartheta)^{-2\ep}
\\[1mm]
&\times
\int_{-1}^{1}\rd(\cos\varphi)
(\sin\varphi)^{-1-2\ep}\,(1-\cos\vartheta)^{-1}
\\[1mm]
&\qquad\times
[2-(1+\cos\chi)\cos\vartheta-\sin\chi\sin\vartheta\cos\varphi]^{-1}
\\[1mm]
&\times
\int_{0}^{y_{0}}\rd y\,y^{-1-2\ep}(1-y)^{d'_{0}}
\,\cJ\bigg(\frac{(1-\cos\vartheta)}{2-y(1+\cos\vartheta)};\ep,y_{0},d'_{0},0
\bigg)
\, ,
\esp
\label{eq:JJirint}
\eeq
and
\beq
\bsp
&
\cJJ{kr}(\Y{i}{k};\ep,y_{0},d'_{0}) =
- 8\Y{i}{k}
\,\frac{\Gamma^{2}(1-\ep)}{2\pi\Gamma(1-2\ep)}
\,\int_{-1}^{1}\rd(\cos\vartheta)\,(\sin\vartheta)^{-2\ep}
\\[1mm]
&\qquad\times
\int_{-1}^{1}\rd(\cos\varphi)\,
(\sin\varphi)^{-1-2\ep}(1-\cos\vartheta)^{-1}
\\[1mm]
&\qquad\times
\int_{0}^{y_{0}}\rd y\,y^{-1-2\ep}(1-y)^{d'_{0}}
\,[2-(1+\cos\chi)\cos\vartheta-\sin\chi\sin\vartheta\cos\varphi]^{-1}
\\[1mm]
&\qquad\qquad\times
\cJ\bigg(\frac{(1-\cos\chi\cos\vartheta-\sin\chi\sin\vartheta\cos\varphi)}
{2-y(1+\cos\chi\cos\vartheta+\sin\chi\sin\vartheta\cos\phi)}
;\ep,y_{0},d'_{0},0
\bigg)
\, ,
\esp
\label{eq:JJkrint}
\eeq
with $\cJ$ given in \eqn{eq:Jint}. The three integrals in \eqnss{eq:JJikint}{eq:JJkrint} need to be 
calculated for $\Y{i}{k} \in [0,1]$ as expansion in $\ep$. 
Explicit results and details of the computation (and values for the parameters $y_{0}$ and $d'_{0}$) 
for these integrals are presented in \sects{sec:Method}{sec:JJints}.

The final case is when the soft integral appears in the argument of a soft-collinear one,
\beq
\bsp
\cKJ{}(\ep,y_{0},d'_{0}) &=
2\,\frac{\Gamma^{2}(1-\ep)}{2\pi\Gamma(1-2\ep)}
\,\int_{-1}^{1}\rd(\cos\vartheta)\,(\sin\vartheta)^{-2\ep}
\\[1mm]
&\times
\int_{-1}^{1}\rd(\cos\varphi)\,(\sin\varphi)^{-1-2\ep}
\,\int_{0}^{y_{0}}\rd y\,y^{-1-2\ep}\,(1-y)^{d'_{0}-1}
\\[1mm]
&\qquad\times
\frac{2-y(1+\cos\vartheta)}{1-\cos\vartheta}
\,\cJ\left(\frac{1-\cos\vartheta}{2-y(1+\cos\vartheta)};\ep,y_0,d'_0,0\right)
\,,
\label{eq:KIint}
\esp
\eeq
which is again independent of the kinematics, i.e. the
coefficients of the expansion in $\ep$ are pure numbers. 
Details of the computation and the parameters $y_{0}$ and $d'_{0}$ are given in \sects{sec:Method}{sec:KIint}.

\section{The method of Mellin-Barnes representations}
\label{sec:Method}

In this Section we briefly review the essential steps in the derivation of 
MB representations for the integrals of \sects{sec:basicintegrals}{sec:iterated}.
The starting point is the well known basic formula,
\beeq
\label{eq:mbformula}
\frac{1}{(a+b)^{\nu}}=\frac{1}{\Gamma(\nu)}\int_{q-\i\infty}^{q+\i\infty}
\frac{\rd z}{2\pi i}\,a^{-\nu-z}\,b^{z}\,\Gamma(\nu+z)\Gamma(-z)\, ,
\eeeq
where $\nu$ and $q$ are real numbers (the case of $\nu = 0$ is trivial) 
and $q$ sets the asymptotic position of the integration contour. 
The application of \eqn{eq:mbformula} to Feynman integral calculus was initiated in 
\Refs{Bergere:1973fq,Usyukina:1975yg} 
(see also \Ref{Smirnov:2004ym}) and is an algorithmic procedure which can be completely automatized, 
as e.g. in the \texttt{Ambre.m} package~\cite{Gluza:2007rt} in \texttt{MATHEMATICA}.

In general, the contour  in \eqn{eq:mbformula} is not necessarily a straight
line and its standard definition is such that the poles of $\Gamma(\nu+z)$ (at
$z = -i -\nu$ with $i$ being non-negative integer) are all to the left
and the poles of $\Gamma(-z)$ (at non-negative integers) are all to the right of it.  
The condition on the poles of the $\Gamma$-functions can be satisfied by such a contour in \eqn{eq:mbformula} 
if and only if $q < 0$ and $\nu > 0$.
However, as a key observation, \Ref{Tausk:1999vh} realized  
straight-line contours parallel to the imaginary axis in an algorithmic way. 
If $\nu < 0$, we start with a curved contour that fulfills the condition on the pole and then deform
it into a straight line taking into account the residua of the crossed poles according to Cauchy's theorem. 
This procedure lends itself to implementation in computer codes for the evaluation and 
manipulation of MB integrals, such as in the \texttt{MB.m} package~\cite{Czakon:2005rk}.

\FIGURE{
\includegraphics[scale=1.25]{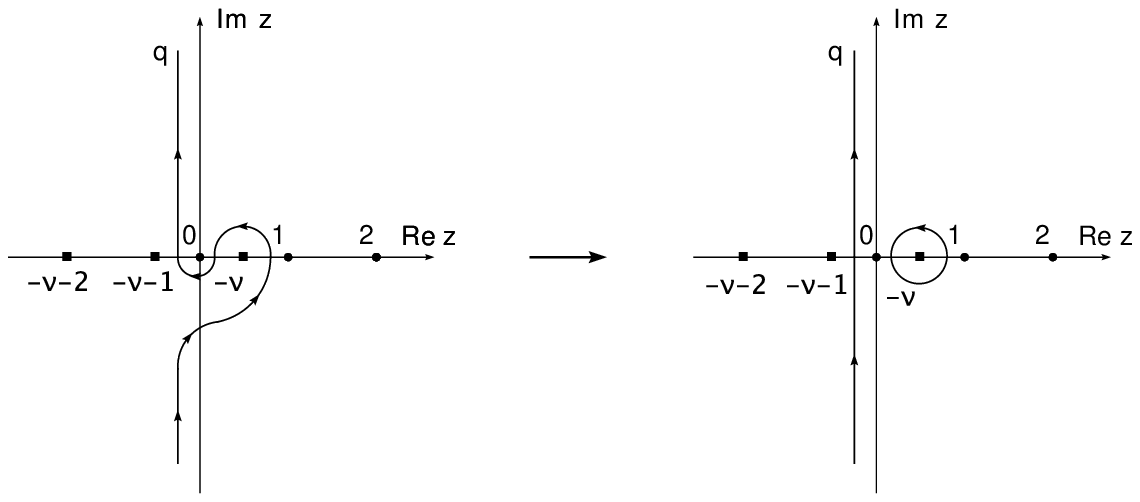}
\caption{
The deformation of a curved contour into the sum of a straight line and a
circle for $\nu=-1/2$ and $q=-1/4$. 
\label{fig:contour}}
}

For instance in \eqn{eq:mbformula}, if $\nu=-1/2$, a possible good choice for $q$ is $q=-1/4$. 
The curved contour depicted in \fig{fig:contour} on the left fulfills the conditions on the
poles of the $\Gamma$-functions. 
Contour deformation results in a straight line and a circle around the pole at $z = -\nu = 1/2$
as shown graphically in \fig{fig:contour} on the right.
Therefore, the MB representation of $(a+b)^{1/2}$ in terms of a
vertical straight line contour is given by 
\beeq
\label{eq:exmbformula}
(a+b)^{1/2}=\frac{1}{\Gamma(-1/2)}\int_{-1/4-\i\infty}^{-1/4+\i\infty}
\frac{\rd z}{2\pi i}\,a^{1/2-z}\,b^{z}\,\Gamma(-1/2+z)\Gamma(-z)+\sqrt{b}
\, ,
\eeeq   
where the first term corresponds to the integral along the straight
line with $q=-1/4$ and the second one to the integral along a circle
surrounding the pole in $z=1/2$ and evaluated according to Cauchy's
theorem.
Note that \eqn{eq:mbformula} is not valid for negative integer values
of $\nu$ because of $\Gamma(\nu)$ in the denominator.
In these cases we use the binomial expansion
\beeq\label{eq:newtonexp}
\frac{1}{(a+b)^{\nu}}=\sum_{i=0}^{-\nu}{-\nu\choose i}a^{i}b^{-\nu-i}\,,
\qquad \textrm{for $\nu$ being negative integer.}
\eeeq

We use \eqns{eq:mbformula}{eq:newtonexp} to convert all sums in the integrands 
of the soft, collinear and iterated integrals of \sect{sec:Subtractions} into products. 
Then, we apply the relations 
\beeq\label{eq:delta}
(1-x)=\int_{0}^{1}\rd y\, y\,\delta(1-x-y)
\, ,
\eeeq
and
\beeq\label{eq:gammaint}
\int_{0}^{1}\prod_{i=1}^{n} \rd x_{i}\, x_{i}^{\,p_{i}-1}
\,\delta\bigg(1-\sum_{j=1}^{n}x_{j}\bigg) =
\prod_{i=1}^{n}\Gamma(p_{i})\big/\Gamma\bigg(\sum_{j=1}^{n}p_{j}\bigg)
\, ,
\eeeq
to obtain a representation of the original integrals in terms of MB
integrals where all the integrations over $\alpha,\,v,\,\cos(\theta),\,\cos(\phi)$ are performed and only 
complex integrations along straight lines parallel to the imaginary axis are left 
following the procedure discussed below \eqn{eq:mbformula}. 
Upon deformation of the curved complex contours all singularities in $\ep$ are
extracted so that it is safe to expand in $\ep$ around zero 
before doing the complex integration.
In this way, the MB representations of the required coefficients of the Laurent expansions 
of the integrals of \sect{sec:Subtractions} are obtained.
In the next step we convert the complex contour integrations into harmonic sums using Cauchy's
theorem and finally we evaluate the sums. 
For the computation of all the harmonic sums we have used algorithms for harmonic and nested sums 
of \Refs{Vermaseren:1998uu,Moch:2001zr} as implemented in the \texttt{XSummer} package \cite{Moch:2005uc}. 
Typically, symbolic summation of single-scale nested sums leads at intermediate stages of the
calculation to harmonic polylogarithms (HPLs) (see \Ref{Remiddi:1999ew} for a definition).
However, in all cases where analytic result have been obtained
by summing series of residues, the HPLs could be converted to standard polylogarithms 
(see also \Ref{Aglietti:2008fe} for a discussion of the class of functions
appearing in the integrated real-virtual counterterms).

As an example let us consider the following integral: 
\beeq \label{eq:example}
\cE(x;\ep,d_0) &=&
x^{2}
\int_0^{1}\! \rd \alpha\, \alpha^{-1-\ep}
\,(1-\alpha)^{2d_0}\,[\alpha+(1-\alpha)x]^{-1-\ep}
\,[2\alpha+(1-\alpha)x]^{-1}
\, ,
\eeeq
which is a typical contribution to the collinear integrals defined in \eqn{eq:Iint}. 
The integral in \eqn{eq:example} is clearly divergent 
in the limit $\ep = 0$ due to the factor $\alpha^{-1-\ep}$ in its integrand. 
The first step is to write the MB representation of the integral in \eqn{eq:example}.  
To do this we use \eqn{eq:mbformula} twice then \eqns{eq:delta}{eq:gammaint}. 
We obtain
\beeq
\label{eq:mbrepepnozero}
\cE(x;\ep,d_0)&=&\int_{q_{1}-\i\infty}^{q_{1}+\i\infty}\frac{\rd z_{1}}{2\pi i}
\int_{q_{2}-\i\infty}^{q_{2}+\i\infty}\frac{\rd z_{2}}{2\pi i}\,
2^{z_{2}}\,x^{-\ep-z_{1}-z_{2}}
\nn\\ &&
\times\Gamma\left(\begin{array}{c}
-z_{1},\, -z_{2},\,2d_{0}-1-\ep-z_{1}-z_{2},\,1+\ep+z_{1},\,1+z_{2},\,-\ep+z_{1}+z_{2}\\
2d_{0}-1-2\ep,\,1+\ep \\
\end{array}\right),
\eeeq
where we have introduced the notation
\beeq
\Gamma\left(\begin{array}{cccc}
a_{1},& a_{2},&\dots,&a_{n}\\
b_{1},& b_{2},&\dots,&b_{m}\\
\end{array}\right)=\prod_{i=1}^{n}\Gamma(a_{i})/\prod_{j=1}^{m}\Gamma(b_{j})
\, .
\eeeq
For $d_{0}\geq 2$ we choose $q_{1}=-1/4$ and $q_{2}=-1/8$ and curved
contours such that the real parts of the arguments of all $\Gamma$-functions 
remain positive on them. 
Note that this implements the requirement that the contour separates 
the left poles from the right ones. 
In order to use straight-line contours, we
must add contributions from two residua: the first is due
to the residue coming from the pole in $z_{2}=\ep-z_{1}$ and then in
the resulting one-dimensional MB integral the second is due to the
residue in $z_{1}=\ep$. 
Adding these contributions to the starting representation of \eqn{eq:mbrepepnozero}, 
we find the MB representation of \eqn{eq:example} with $\ep$ close to zero to be given by:
\beeq
\label{eq:mbrepepzero}
\cE(x;\ep,d_0) &=& x^{-2\ep}\Gamma\left(\begin{array}{c}
-\ep,\, 1+2\ep\\
1+\ep\\
\end{array}\right)
\nn\\[2mm]
&&+
\int_{q_{1}-\i\infty}^{q_{1}+\i\infty}\frac{\rd z_{1}}{2\pi i}\,2^{\ep-z_{1}}\,x^{-2\ep}
\Gamma\left(\begin{array}{c}
-z_{1},\, -\ep+z_{1},\,1+\ep+z_{1},\,1+\ep-z_{1}\\
1+\ep\\
\end{array}\right)
\nn\\[2mm]
&&+\int_{q_{1}-\i\infty}^{q_{1}+\i\infty}\frac{\rd z_{1}}{2\pi i}
\int_{q_{2}-\i\infty}^{q_{2}+\i\infty}\frac{\rd z_{2}}{2\pi i}\,
2^{z_{2}}\,x^{-\ep-z_{1}-z_{2}}
\nn\\[2mm]
&&\times\Gamma\left(\begin{array}{c}
-z_{1},\, -z_{2},\,2d_{0}-1-\ep-z_{1}-z_{2},\,1+\ep+z_{1},\,1+z_{2},\,-\ep+z_{1}+z_{2}\\
2d_{0}-1-2\ep,\,1+\ep \label{cubaevaluation}\\
\end{array}\right)
\, .
\eeeq
At this point we see that the singularity in $\ep=0$ is isolated in
the first term of this equation. In particular the pole comes from the
factor $\Gamma(-\ep)$ of this term. This shows that the extraction of
poles comes out in a very convenient way: in practice we have only
deformed contours and computed residua. 
As a matter of fact, this is one of the strong points in the application of MB methods to
phase space integrals, and the straightforward way of extracting infrared poles 
has been already discussed in \Refs{Gluza:2007uw,Gluza:2008tk}.

As a next step we can perform the expansion around $\ep=0$ and we obtain
\beeq
\cE(x;\ep,d_0) &=&-\frac{1}{\ep}+2\log(x)+\int_{q_{1}-\i\infty}^{q_{1}+\i\infty}
\frac{\rd z_{1}}{2\pi i}\, 2^{-z_{1}}\Gamma(1-z_{1},\,-z_{1},\,z_{1},\,1+z_{1})
\nonumber\\[2mm] &&+
\int_{q_{1}-\i\infty}^{q_{1}+\i\infty}
\frac{\rd z_{1}}{2\pi i}\int_{q_{2}-\i\infty}^{q_{2}+\i\infty}
\frac{\rd z_{2}}{2\pi i}\,2^{z_{2}}\,x^{-z_{1}-z_{2}}
\nonumber\\[2mm] &&
\times\Gamma\left(\begin{array}{c}
-z_{1},\, -z_{2},\,2d_{0}-1-z_{1}-z_{2},\,1+z_{1},\,1+z_{2},\,z_{1}+z_{2}\\
2d_{0}-1\\
\end{array}\right)\, .
\eeeq
The first integral can be easily computed. We close the contour to the
right and compute the residua coming from the poles enclosed in it at
$z_{1}=n;\, n\geq 0$. The residua are given by $(1/2)^{n}\log(2)$.
Thus, multiplying by an overall minus sign due to the clockwise
orientation of the contour, we find 
\beeq
 \int_{q_{1}-\i\infty}^{q_{1}+\i\infty}
\frac{\rd z_{1}}{2\pi i}\, 2^{-z_{1}}\Gamma(1-z_{1},\,-z_{1},\,z_{1},\,1+z_{1})
=-\sum_{n=0}^{\infty}(1/2)^{n}\log(2)=-2\, \log(2)
\, .
\eeeq
Next we evaluate the second integral closing both contours to the left.
We begin with the integration over the variable $z_{1}$ and we have
two different $\Gamma$-functions that contribute with poles. 
The first one is $\Gamma(1+z_{1})$ which exhibits poles in
$z_{1}=-n;\,n\geq 1$ and the second one is $\Gamma(z_{1}+z_{2})$ which
contributes with poles in $z_{1}=-n-z_{2};\,n\geq 1$. 
Computing these residua, we obtain for the second integral 
\beeq
&&\int_{q_{1}-\i\infty}^{q_{1}+\i\infty}
\frac{\rd z_{1}}{2\pi i}\int_{q_{2}-\i\infty}^{q_{2}+\i\infty}
\frac{\rd z_{2}}{2\pi i}\,2^{z_{2}}\,x^{-z_{1}-z_{2}}
\Gamma\left(\begin{array}{c}
-z_{1},\, -z_{2},\,2d_{0}-1-z_{1}-z_{2},\,1+z_{1},\,1+z_{2},\,z_{1}+z_{2}\\
2d_{0}-1\\
\end{array}\right)
\nn
\\[2mm] &&\qquad
=\sum_{n=1}^{\infty}\int_{q_{2}-\i\infty}^{q_{2}+\i\infty}\frac{\rd z_{2}}{2\pi i}
\Bigg[ \frac{(-1)^{n+1}}{(n-1)!}\,2^{z_2}\,x^{n-z_{2}}\,\Gamma
\left(\begin{array}{c}
n,\, -z_{2},\,2d_{0}-1+n-z_{2},\,1+z_{2},\,-n+z_{2}\\
2d_{0}-1\\
\end{array}\right)
\nn\\[2mm] &&\qquad\qquad\qquad\qquad\qquad
+\frac{(-x)^{n}}{n!}\,2^{z_2}\,\Gamma
\left(\begin{array}{c}
2d_{0}-1+n,\,1-n-z_{2},\,-z_{2},\,1+z_{2},\,n+z_{2}\\
2d_{0}-1\\
\end{array}\right)\Bigg].
\eeeq 
Now we can do the remaining integration over $z_{2}$. 
In this case the poles of both the integrands are in $z_{2}=-m;\, m\geq 1$ and
the corresponding residua are given by
\beeq
\label{eq:ressums}
\cE(x;\ep,d_0) &=&-\frac{1}{\ep}+2\log\left(\frac{x}{2}\right)-
\log(2)\sum_{m,n=1}^{\infty}\left(\frac{1}{2}\right)^{m}x^{n}
{2d_{0}-2+n\choose n}
\\[2mm]\nn
&&-\sum_{m,n=1}^{\infty}\left(\frac{x}{2}\right)^{m}x^{n}
{2d_{0}-2+m+n\choose m+n}\left[S_{1}(2d_{0}-2+m+n)-S_{1}(m+n)+
\log\left(\frac{x}{2}\right)\right],
\eeeq
where the harmonic sums $S_{1}(n)$ are defined as~\cite{Vermaseren:1998uu,Moch:2001zr}
\beeq
S_{1}(n)=\sum_{i=1}^{n}\frac{1}{i}=\psi(n+1)+\gamma_{E}
\, ,
\eeeq
with $\gamma_{E}$ being Euler's constant and $\psi(x)$ being the polygamma function, 
i.e. the first derivative of the logarithm of the $\Gamma$-function. 
The first double sum amounts to 
\beeq
\sum_{m,n=1}^{\infty}\left(\frac{1}{2}\right)^{m}x^{n}
{2d_{0}-2+n\choose n}=
-\left(1-\frac{1}{(1-x)^{2d_{0}-1}}\right) 
\, .
\eeeq
We are not able to perform the second summation for arbitrary $d_0$.
However, choosing integer values, $d_{0}\geq 2$, these sums simplify significantly.
Indeed, if $d_{0}$ is a positive integer, then both
\beeq
{2d_{0}-2+m+n\choose m+n}
\, , \qquad 
{2d_{0}-2+m+n\choose m+n}\left[S_{1}(2d_{0}-2+m+n)-S_{1}(m+n)\right]
\, ,
\eeeq 
are polynomials in $m$ and $n$. 
This implies that the double sums in \eqn{eq:ressums} can be written in terms of the functions
\beq
\sum_{n=1}^{\infty}\frac{x^{n}}{n^{k}}=\left\{\begin{array}{cc}
\textrm{Li}_{k}(x)
\qquad &\textrm{if $k\geq 0$} 
\, ,
\\
\\
\frac{1}{(1-x)^{1-k}}\sum_{i=0}^{-k-1}\Big\langle 
\begin{array}{c}-k
\\ i 
\end{array}
\Big\rangle x^{-k-i} 
\qquad&\textrm{if $k< 0$}
\, ,
\end{array}\right.
\eeq
where $\textrm{Li}_{k}(x)$ are the classical polylogarithms \cite{Lewin:1981}
and $\Big\langle \begin{array}{c}-k\\ i \end{array}\Big\rangle$ are the
Eulerian numbers:
\beeq
\Big\langle \begin{array}{c}-k\\ i 
\end{array}\Big\rangle=
\sum_{j=0}^{i+1}(-1)^{j}{-k+1\choose j}(i-j+1)^{-k};\qquad k< 0.
\eeeq
Therefore, \eqn{eq:ressums} becomes for example with the choice $d_{0}=2$, 
\beeq
\cE(x;\ep,d_0) &=&-\frac{1}{\ep}+
\log(2)\left(1-\frac{1}{(1-x)^{3}}\right)-\frac{x^{2}(3x^{2}-15x+14)}{2(1-x)^{2}(2-x)^2}
\nonumber\\[2mm] &&
+\frac{(x^6-9x^{5}+33x^{4}-78x^{3}+108x^{2}-72x+16)}{(1-x)^{3}(2-x)^{3}}\, \log
\left(\frac{x}{2}\right)+\O(\ep)
\, .
\label{eq:finalresex}
\eeeq
Looking at this expression we notice that even if the integral in \eqn{eq:example} 
is well defined for $x=1$ or $x=2$
some of its individual terms diverge in these limits.
Nevertheless, the full result has a well defined limit in $x=1$ or $x=2$. 
Indeed,
\beeq
\lim_{x\to 1}\cE(x;\ep,d_0)&=&-\frac{1}{\ep}+\frac{53}{6}-16\log(2)+\O(\ep)
\,,
\\[2mm]
\lim_{x\to 2}\cE(x;\ep,d_0)&=&-\frac{1}{\ep}-\frac{8}{3}+2\log(2)+\O(\ep)
\,.
\eeeq

This completes the discussion of our example and demonstrates that the Laurent coefficients 
are given by simple functions in $x$ only.  
Looking back at how \eqn{eq:mbformula} has enabled us to arrive at \eqn{eq:mbrepepnozero} 
starting from \eqn{eq:example} it is obvious that more complicated
integrals such as nested ones defined in \sect{sec:iterated} 
result in increased numbers of Mellin integrations and shifted arguments 
of the $\Gamma$-functions. 
Also, in the case of nested integrals the order of the singularities is higher. 
However, the extraction of the poles in $\ep$ always reduces the dimensionality of
the MB integrals (in our example from two to zero).
Hence, in general if we start from a high-dimensional MB representation,
the contributions to the poles' coefficients have a much lower
dimensionality of the Mellin integrals, which allows for an 
analytic computation of the coefficients of the poles in the $\ep$
expansion even for the most complicated integrals.
This example also shows that for the analytic computation
by means of a MB representation  one should choose $d_0,d_0'$ to be positive integers
and transform the regions of integrations in the integrals defined in \sect{sec:Subtractions}
to $[0,1]$.
In this work we simply choose $\alpha_0=y_0=1$ and consider the cases $d_0=d_0'=2,3$ which as discussed in \Ref{Somogyi:2008fc} 
are the natural choices for the infrared subtraction for processes with two and three outgoing jets respectively. 
Nevertheless, we
stress that in principle {\em any} choice of $d_0,d'_0 \geq 2$ can be used in a
computation of $m$-jet production, for {\em any} $m$. Thus there is no need
to recompute any integrals even for processes with more than three jets.
Furthermore, the appearance of a hypergeometric function in the integrand of
\eqn{eq:example} as happens for example in the last row of
\tab{tab:I1ints} does not essentially change the complexity of the computation.  
The reason is that the hypergeometric function $_{2}F_{1}$ has a simple MB representation:

\beeq
_{2}F_{1}(a,b,c;w)=\int_{q-\i\infty}^{q+\i\infty}
\frac{\rd z}{2\pi i}\,(-w)^{z}\, \Gamma\left(\begin{array}{c}
c,\,a+z,\, b+z,\,-z\\
a,\,b,\,c+z\\
\end{array}\right)
\,,
\eeeq
where the integration contour separates the poles of the $\Gamma(\dots +z)$ 
functions from the poles of the $\Gamma(\dots -z)$ function as usual. 
 
In closing this Section, we would like to mention another virtue of the MB method.
For a given phase space integral of \sect{sec:Subtractions}, the corresponding
MB representations show good convergence properties if evaluated numerically
along the complex contours.
Thus, the multidimensional numerical integration of MB integrals, such as in 
\eqn{cubaevaluation} is straightforward with the help 
of the \texttt{CUBA} library~\cite{Hahn:2004fe}, which 
provides an independent check. Moreover, it also presents a quick and 
reliable way of obtaining 
numerical results for the (smooth) $\O(\ep^0)$ terms in the Laurent expansions
of all integrals for the real-virtual counterterms in the paper.
 

\section{Collinear integrals $\cI$}
\label{sec:Iints}

In this Section, we show the analytic results for the collinear
integrals defined in \eqn{eq:Iint} for which the case $\kappa=0$ is
needed only for the first row in \tab{tab:I1ints} and $\kappa=1$
is needed for all of them. Analytic expressions for the first two cases
of \tab{tab:I1ints} have already been computed in
\Ref{Aglietti:2008fe}. Here we fix $d_0=3$ and give the explicit
expressions for this case as an illustration of the form of our results. 
For the Laurent expansion we obtain
\beeq
\cI(x;\ep;\alpha_0=1,d_0;\kappa,k,\delta,g_I^{(\pm )})&=&
\frac{\delta_{k,-1}}{2(2-\delta)}\frac{1}{\ep^{2}}-
\left[\frac{2\delta_{k,-1}\log(x)}{3-\delta}+\frac{1-\delta_{k,-1}}{2[1+k(1-\delta_{k,-1})]}\right] 
\frac{1}{\ep}
\nn\\[2mm] &&
+\, \delta_{\kappa,1}\,\cG_{I,k}^{(\pm)}(x)+\cF(x;\ep,d_0,k)+\O(\ep),
\eeeq
where $I=A,B,C,D$ (see \tab{tab:I1ints}), $k=-1,0,1,2$ and $\delta_{i,j}$ is
the usual Kronecker $\delta$.  
Here we have introduced the two functions $\cG_{I,k}^{(\pm)}(x)$ and $\cF(x;\ep,d_{0},k)$.
The function $\cG_{I,k}^{(\pm)}$ is a matrix in $I$ (rows) and $k$ (columns) defined as follows:
\beeq
\cG_{I,k}^{(\pm)}(x)=\left(\begin{array}{cccc}
\frac{2}{3}\zeta_{2}+\frac{1}{3}\log^2(x)\,\,\,&1\,\,\,&\frac{1}{2}\,\,\,&\frac{13}{36}\\
\\
\left(\frac{5}{8}\pm\frac{5}{8}\right)\zeta_{2}+\left(\frac{1}{2}\mp\frac{1}{2}\right)\log^2(x)\,\,\,&1\,\,\,&
\frac{1}{2}\pm\frac{1}{4}\,\,\,&\frac{13}{16}\pm\frac{1}{4}\\
\\
\left(\frac{2}{3}\pm\frac{1}{2}\right)\zeta_{2}+\frac{1}{3}\log^{2}(x)\,\,\,&1\pm\frac{1}{2}\,\,\,&
\frac{1}{2}\pm\frac{3}{8}\,\,\,&\frac{13}{36}\pm\frac{11}{36}\\
\\
\left(\frac{13}{36}\mp\frac{1}{16}\right)\zeta_{2}+\left(\frac{1}{2}\pm\frac{1}{2}\right)\log^{2}(x)\,\,\,&
1\pm\frac{1}{2}\,\,\,&\frac{1}{2}\pm\frac{1}{8}\,\,\,&\frac{13}{36}\pm \frac{1}{18}
\end{array}\right),
\eeeq
and choosing e.g. $d_0=3$ for the function $\cF(x;\ep,d_{0},-1)$, we obtain
\beeq 
\label{eq:prima}
\cF(x;\ep,d_{0}=3,-1) &=& 
-\frac{3}{2}\zeta_{2}+\log^{2}(x) - \frac{1}{24} P_{0,-1}^{(5)} (x; 35,-133,188,-116,0,0)
\\
&& 
- \frac{1}{12} P_{1,-1}^{(5)} (x; 25,-116,212,-192,96,0) - P_{2,-1}^{(5)} (x; 1,-5,10,-10,5,2)
\,,
\nn
\\
\cF(x;\ep,d_{0}=3,0) &=& 
- \frac{1}{12}P_{0,0}^{(5)}(x; 49,-193,281,-173,24,0)
+ P_{1,0}^{(5)} (x; 1,-5,10,-10,5,2)
\,,
\\
\cF(x;\ep,d_{0}=3,1) &=& 
\frac{1}{2}\,\cF(x;\ep,d_{0}=3,0)
\,,
\\
\label{eq:ultima}
\cF(x;\ep,d_{0}=3,2) &=&
  \frac{80 x}{3 (2 - x)^6}\, \log(2)
+ \frac{(1 - x)^6}{36 (2 - x)^6}
  P_{0,2}^{(11)}
(x;51,-861,6523,-29212,
\nn\\
&&\qquad
85505,-171607,241761,-240096,164864,-74000,19120,-2048)
\nn\\
&&
+ \frac{(1 - x)^6}{3 (2 - x)^6} P_{1,2}^{(11)}(x;1,-17,130,-590,1765,-3734,5748,-6360,
\nn\\
&&\qquad
4880,-2480,784,-128)    
\,.    
\eeeq
Here we introduced the short-hand notation
\beq
P_{n,k}^{(m)}(x; a_m^{(k)},\dots, a_0^{(k)})
= \frac{\Li_n(1-x)}{(1-x)^m} \sum_{i=0}^m a_i^{(k)} x^i
\,.
\eeq
According to their definition, the limit of the functions 
given in \eqnss{eq:prima}{eq:ultima} must be finite
in $x=1$ even if some terms are separately divergent.  
Indeed computing the limit at $x=1$ we find
\begin{eqnarray}
\lim_{x\to 1}\cF(x;\ep,d_{0}=3,-1)&=&-\frac{8731}{3600}-\frac{3}{2}\, \zeta_{2}\,,\\
\lim_{x\to 1}\cF(x;\ep,d_{0}=3,0)&=&-\frac{257}{60}\,,\\
\lim_{x\to 1}\cF(x;\ep,d_{0}=3,1)&=&-\frac{257}{120}\,,\\
\lim_{x\to 1}\cF(x;\ep,d_{0}=3,2)&=&-\frac{1801}{90}+\frac{80}{3}\log(2)\,.
\end{eqnarray} 

In \fig{fig:CDfigs} we compare the analytic and numeric results for 
the $\ep^0$ coefficient in the expansion of $\cI(x,\ep;1,3;1,-1,0,g_C^{(+)})$ 
and $\cI(x,\ep;1,3;1,-1,1,g_D^{(+)})$  for $k=-1$,  $\alpha_0=1$ and $d_0=3$
as representative examples. The agreement between the two computations 
is excellent for the whole $x$-range. The numeric results have been obtained
using standard residuum subtraction and a Monte Carlo integration program as
explained in detail in \Refs{Somogyi:2008fc,Aglietti:2008fe}. This shows that 
the expansion
coefficients of all the collinear integrals $\cI{}$ and hence also of the
collinear subtraction terms are smooth functions of the kinematical variable $x$. 

\FIGURE{
\includegraphics[scale=0.72]{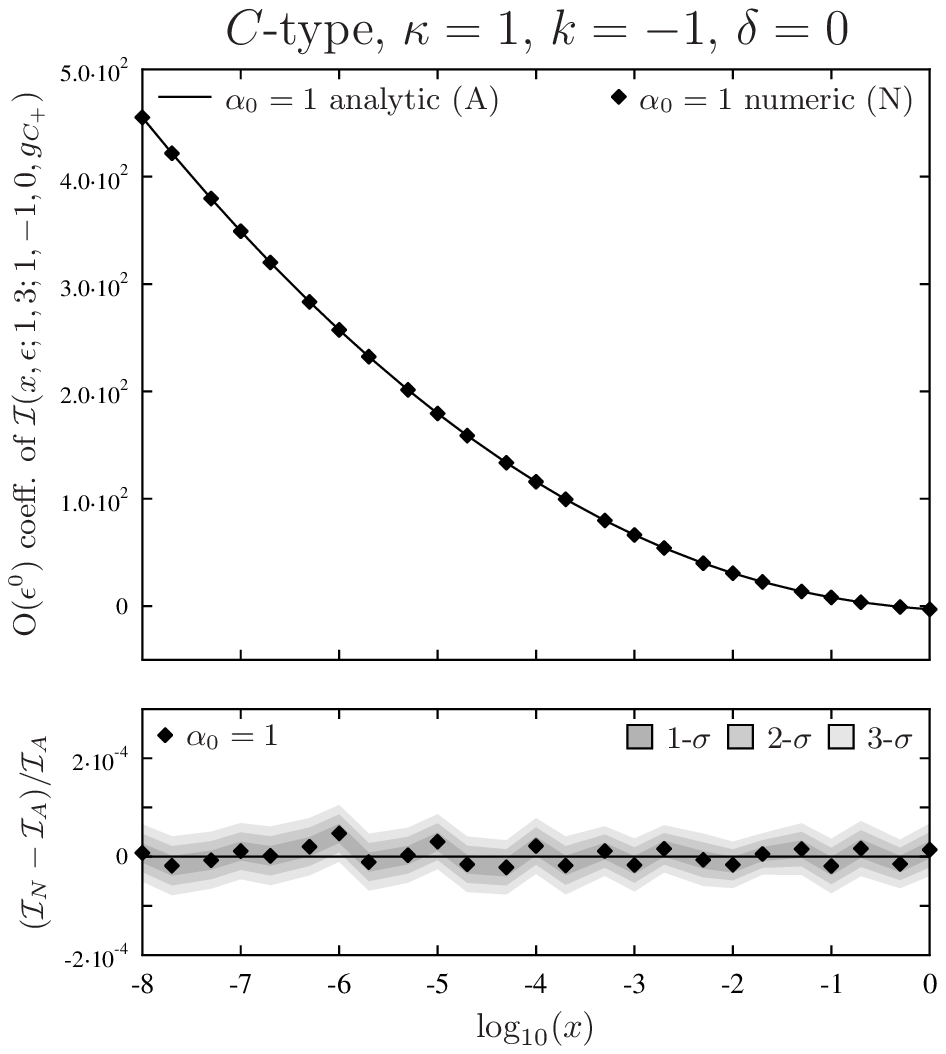}
\includegraphics[scale=0.72]{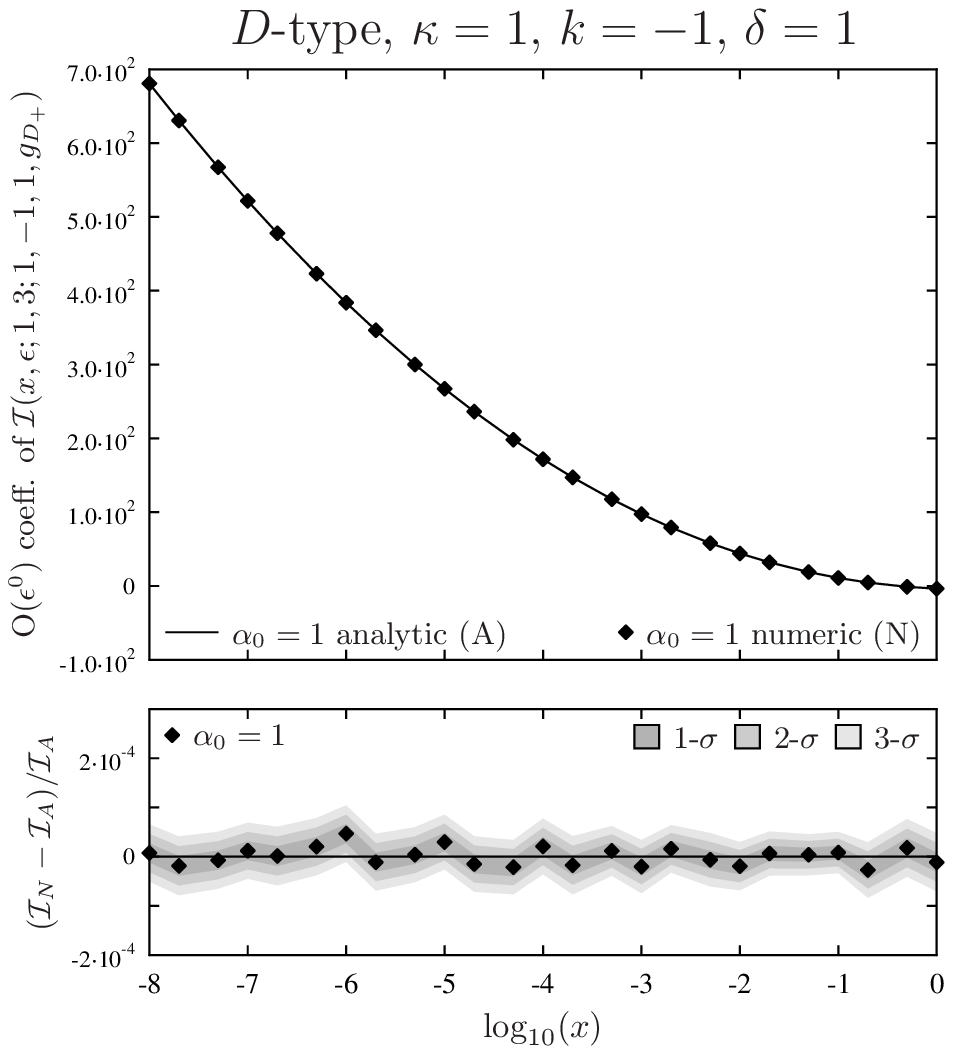}
\caption{\label{fig:CDfigs}
Representative results for the $C$-type and $D$-type integrals. 
The plots show the coefficient of the $\Oe{0}$ term for $k=-1$ in 
$\cI(x,\ep;1,3;1,-1,0,g_C^{(+)})$ (left) and $\cI(x,\ep;1,3;1,-1,1,g_D^{(+)})$ 
(right) with $d_0=3$ and $\alpha_0=1$.}
}

The complete results for all necessary cases (like in the later Sections) 
are of considerable size, such that we shall not list them here.  
They are all contained in a \texttt{MATHEMATICA} file provided with the sources 
of the paper on the archive {\tt http://arXiv.org}.


\section{Nested collinear-type $\cII{}$ and $\cIJ{}$ integrals}
\label{sec:IIints}

In this Section we discuss the analytic computation of the nested collinear 
integrals defined in \eqnss{eq:IIiint}{eq:IJint}. 

As an example we show explicitly the fully analytic result for the case
$\cII{r}(x,\ep;1,3;-1,2)$ for which we were able to compute the complete pole structure analytically.
Choosing $d_0=3$ and $\alpha_0=1$ we get 
\begin{eqnarray}
&&\cII{r}(x,\ep;1,3;-1,2)=-\frac{1}{12}\,\frac{1}{\ep^3}+\left(-\frac{2}{9}+\frac{1}{3}\log(x)\right)\,\frac{1}{\ep^2}
+\bigg[\frac{1}{(1 - x)^5}\,\bigg(-\frac{1}{3}\,\zeta_2-\frac{25}{36}\,\log(x)\nonumber\\
&&\quad
+\frac{1}{3}\,\log(1 - x)\log(x)+\frac{1}{3}\,\textrm{Li}_{2}(x)\bigg)+ 
\frac{1}{(1-x/2)^5}\,\left(\frac{1}{6}\log\left(\frac{x}{2}\right)\right)+\frac{1}{(1 - x)^4}\,\left(-\frac{13}{36}+\frac{1}{6}\log(x)\right)\nonumber\\
&&\quad
+\frac{1/6}{(1-x/2)^4}+\frac{1}{(1 - x)^3}\,\left(-\frac{7}{72}-\frac{1}{18}\,\log(x)\right)+\frac{1/12}{(1-x/2)^3}\nonumber\\ 
&&\quad
+\frac{1}{(1 - x)^2}\,\left(-\frac{1}{6}-\frac{2}{9}\,\log(x)\right)+\frac{1/18}{(1-x/2)^2}+\frac{1}{(1 - x)}\,\left(-\frac{25}{72}-\frac{7}{12}\,\log(x)\right)\nonumber\\
&&\quad
+\frac{1/24}{(1-x/2)}+\frac{31}{216}+\frac{1}{6}\,\log(2)+\frac{19}{9}\,\log(x)+\frac{2}{3}\,
\log(1 - x) \log(x)-\frac{2}{3}\log^2(x)\nonumber\\
&&\quad
+\frac{2}{3}\,\textrm{Li}_{2}(x)\bigg]\,\frac{1}{\ep}+\O(\ep^0). \label{IIexplicit}
\end{eqnarray}
This result is representative, because its form is typical of all the collinear nested integrals.
The plot of the $\O(\ep^{-1})$ coefficient of this Laurent expansion for 
the integral $\cII{r}(x,\ep;1,3;-1,2)$ is shown on the right side of
\fig{fig:IIfigs} together with the comparison with the numerical evaluation obtained using
standard residuum subtraction and Monte Carlo numerical integration. 
On the left side of \fig{fig:IIfigs} we plot the same coefficient of the Laurent
expansion for the integral $\cII{i}(x,\ep;1,3;-1,2)$. 
For both cases we note that the agreement between the numerical evaluation and the
analytic result is excellent. These plots show also that the coefficients of the Laurent expansion
of the nested collinear integrals $\cII{}$ are very smooth functions of $x$.

\FIGURE{
\includegraphics[scale=0.72]{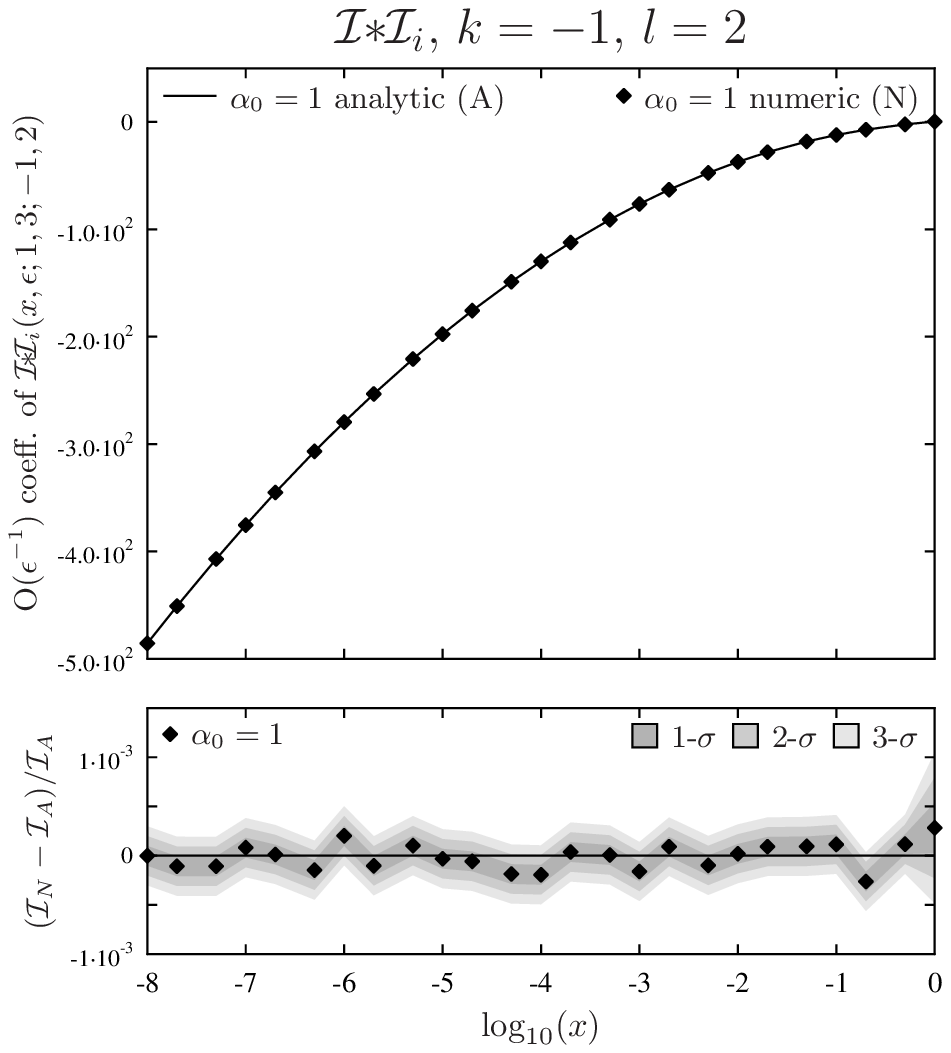}
\includegraphics[scale=0.72]{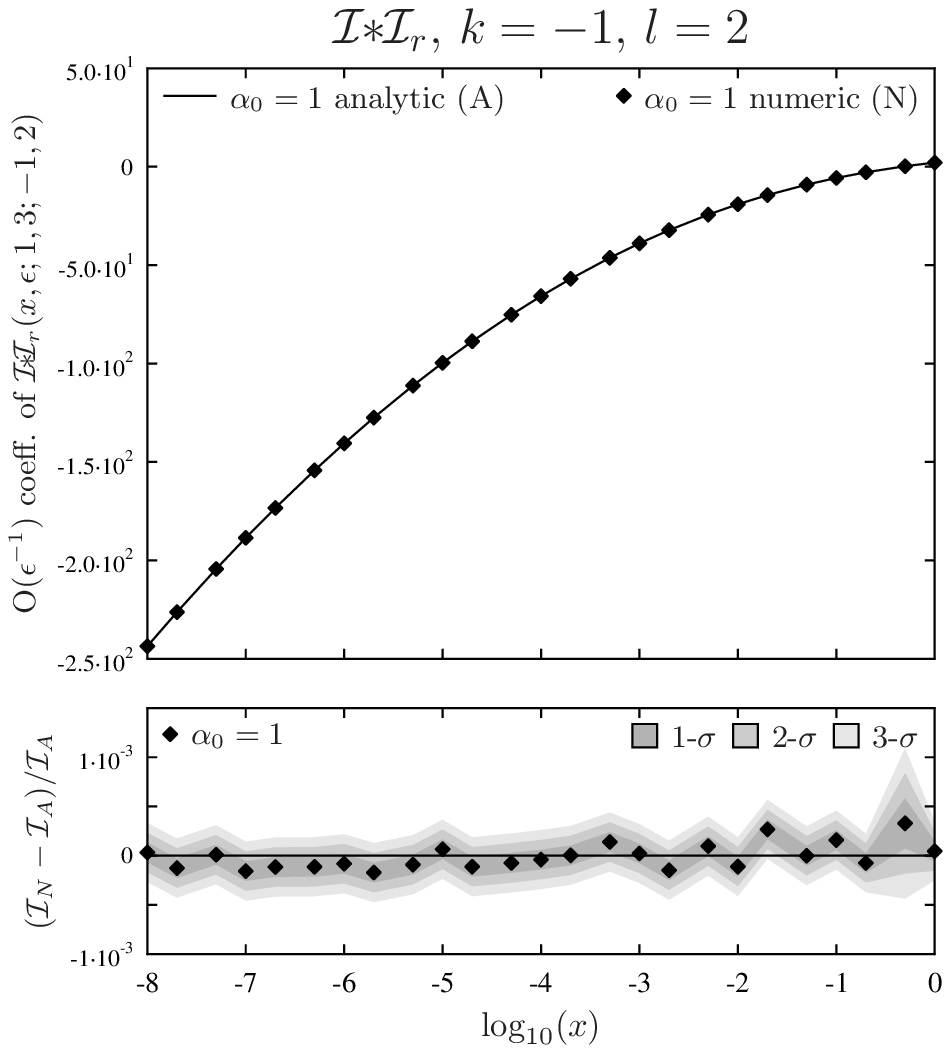}
\vspace{-2em}
\caption{\label{fig:IIfigs}
Representative results for the ${\cII{}}$-type integrals. 
The plots show the coefficient of the $\Oe{-1}$ term for $k=-1$ and $l=2$ in 
$\cII{i}(x,\ep;1,3;-1,2)$ (left) and $\cII{r}(x,\ep;1,3;-1,2)$ 
(right) with $d_0=3$ and $\alpha_0=1$.}
}

We note that in the Laurent expansion
of $\cII{r}(x,\ep;1,3;-1,2)$ in \eqn{IIexplicit} there are some terms that are divergent in
$x=1$. However according to its definition in \eqn{eq:IIrint} the limit in $x=1$ must be
finite. To verify this is a further check of the correctness of the result. 
For the case of \eqn{IIexplicit} we obtain that:
\begin{equation}
\lim_{x\to 1}\cII{r}(x,\ep;1,3;-1,2)=-\frac{1}{12}\,\frac{1}{\ep^3}-\frac{2}{9}\,\frac{1}{\ep^2}+
\left(\frac{3091}{675}+\frac{2}{3}\,\zeta_2-\frac{31}{6}\,\log(2)\right)\,\frac{1}{\ep}+\O(\ep^0).
\end{equation}

The case of $\cII{r}(x,\ep;1,3;2,-1)$ is more difficult. For this integral we are unable to
compute the coefficients of the $\ep$ poles in a fully analytic form. 
The reason is that in its Mellin-Barnes representation
also three-fold MB integrals are involved. For this case the coefficient $\O(\ep^{-3})$ and 
$\O(\ep^{-2})$ are fully analytic but the coefficient of $\O(\ep^{-1})$ is semi-analytic. This last 
coefficient is thus written in terms of an analytic expression to which a
three-fold MB integral must be added. 
The remaining MB integral can be efficiently computed in \texttt{MATHEMATICA} by use of the package
\texttt{MB.m} \cite{Czakon:2005rk}.
Explicitely for $\cII{r}(x,\ep;1,3;2,-1)$ we have:
\begin{eqnarray}
&&\cII{r}(x,\ep;1,3 ;2,-1)=-\frac{1}{6}\,\frac{1}{\ep^3}+\bigg[\frac{1}{(1-x/2)^6}\,\left(-\frac{5}{12}\,\log\left(\frac{x}{2}\right)\right)+ 
\frac{1}{(1 - x)^5}\,\left(\frac{1}{6}\,\log(x)\right)\nonumber\\
&&\quad+\frac{1}{(1-x/2)^5}\,\left(-\frac{5}{12}+\frac{5}{12}\,\log\left(\frac{x}{2}\right)\right)+ 
\frac{1/6}{(1 - x)^4}+\frac{5/24}{(1-x/2)^4}+\frac{1/12}{(1 - x)^3}\nonumber\\
&&\quad+\frac{5/72}{(1-x/2)^3}+\frac{1/18}{(1 - x)^2}+\frac{5/144}{(1-x/2)^2}
+\frac{1/24}{(1-x)}+\frac{1/48}{(1-x/2)}-\frac{59}{72}+\frac{1}{2}\,\log(x)\bigg]\,\frac{1}{\ep^2}
\nonumber\\
&&\quad+\bigg[ 
\frac{1}{(1-x/2)^6}\bigg(\frac{25}{24}\,\zeta_2-\frac{21}{8}\,\log(2)+\frac{5}{4}\,\log^2(2)
+\frac{5}{3}\,\log(2)\,\log(1-x/2)+\frac{21}{8}\,\log(x)\nonumber\\
&&\quad-\frac{5}{2}\,\log(2)\,\log(x)+\frac{5}{12}\,\log(1 - x)\,\log(x)
-\frac{5}{3}\,\log(1 - x/2)\,\log(x)+\frac{5}{4}\,\log^2(x)\nonumber\\
&&\quad-\frac{5}{3}\,\textrm{Li}_2\left(\frac{x}{2}\right)+\frac{5}{12}\,\textrm{Li}_2(x)\bigg)
+\frac{1}{(1-x)^5}\,\bigg(-\frac{1}{3}\,\zeta_2
-\frac{1}{6}\,\log^2(2)-\frac{1}{3}\,\log(2)\,\log(1-x/2)\nonumber\\
&&\quad
+\frac{17}{24}\,\log(x)+\frac{1}{3}\,\log(2)\,\log(x)+\frac{1}{6}\,\log(1-x)\,\log(x)
+\frac{1}{3}\,\log(1-x/2)\,\log(x)\,\nonumber\\
&&\quad-\frac{1}{2}\,\log^2(x)+\frac{1}{3}\textrm{Li}_2\left(\frac{x}{2}\right)
+\frac{1}{6}\,\textrm{Li}_2(x)\bigg)+\frac{1}{(1-x/2)^5}\,\bigg(\frac{23}{24}
-\frac{25}{24}\,\zeta_2+\frac{71}{24}\,\log(2)\nonumber
\\
&&\quad-\frac{5}{4}\,\log^2(2)-\frac{5}{3}\,\log(2)\,\log(1 - x/2)-\frac{17}{8}\,\log(x)
+\frac{5}{2}\,\log(2)\,\log(x)+\nonumber
\end{eqnarray}
\begin{eqnarray}
&&\quad-\frac{5}{12}\,\log(1 - x)\,\log(x)+ 
\frac{5}{3}\,\log\left(1-\frac{x}{2}\right)\,\log(x)-\frac{5}{4}\,\log^2(x)+ 
\frac{5}{3}\,\textrm{Li}_2\left(\frac{x}{2}\right)-\frac{5}{12}\,\textrm{Li}_2(x)\bigg)\nonumber\\
&&\quad+\frac{1}{(1 - x)^4}\,\bigg(
\frac{7}{8}-\frac{1}{4}\,\zeta_2+\frac{2}{3}\,\log(2)-\frac{11}{8}\,\log(x)+ 
\frac{1}{4}\,\log(1 - x)\,\log(x)+\frac{1}{4}\,\textrm{Li}_2(x)\bigg)\nonumber\\
&&\quad+\frac{1}{(1-x/2)^4}\,\left(-\frac{59}{48}+\frac{1}{6}\,\log\left(\frac{x}{2}\right)\right)+ 
\frac{1}{(1 - x)^3}\,\bigg(-\frac{1}{16}
-\frac{1}{12}\,\zeta_2-\frac{7}{36}\,\log(x)\nonumber\\
&&\quad+\frac{1}{12}\,\log(1 - x)\,\log(x)+\frac{1}{12}\,\textrm{Li}_2(x)\bigg)+ 
\frac{1}{(1-x/2)^3}\,\left(-\frac{29}{432}-\frac{1}{6}\,\log(2)+\frac{4}{9}\,\log(x)\right)\nonumber\\
&&\quad+\frac{1}{(1 - x)^2}\,\left(-\frac{1}{27}+\frac{2}{9}\,\log(2)-\frac{23}{36}\,\log(x)\right)+ 
\frac{1}{(1-x/2)^2}\,\left(\frac{211}{864}-\frac{2}{9}\,\log(2)+\frac{7}{9}\,\log(x)\right)\nonumber\\ 
&&\quad+\frac{1}{(1 - x)}\,\left(-\frac{31}{72}-\frac{59}{24}\,\log(x)\right)+ 
\frac{1}{(1-x/2)}\,\left(\frac{139}{288}-\frac{1}{3}\,\log(2)+\frac{4}{3}\,\log(x)\right)
-\frac{1177}{432}+\frac{7}{8}\,\zeta_2\nonumber\\
&&\quad-\,\frac{1}{3}\log(2)+\frac{1}{6}\log^2(2)+\frac{1}{3}\,\log(2)\,\log\left(1 - \frac{x}{2}\right)+
\frac{71}{24}\,\log(x)-\frac{1}{3}\,\log(2)\,\log(x)\nonumber\\
&&\quad+\frac{1}{2}\,\log(1-x)\,\log(x)-\frac{1}{3}\,\log\left(1-\frac{x}{2}\right)\,\log(x)
-\frac{5}{6}\,\log^2(x))-\frac{1}{3}\,\textrm{Li}_2\left(\frac{x}{2}\right)+\frac{1}{2}\,\textrm{Li}_2(x)\nonumber\\
&&+\textrm{MBint}[x]\bigg]\,\frac{1}{\ep}+\O(\ep^0), \label{secondexample}
\end{eqnarray}
where $\textrm{MBint}[x]$ is a three-fold Mellin-Barnes integral, which for this case is given by
\begin{eqnarray}
&&\textrm{MBint}[x]=\int_{q_{1}-\i\infty}^{q_{1}+\i\infty}\frac{\rd z_{1}}{2\pi i}
\int_{q_{2}-\i\infty}^{q_{2}+\i\infty}\frac{\rd z_{2}}{2\pi i}\int_{q_{3}-\i\infty}^{q_{3}+\i\infty}\frac{\rd z_{3}}{2\pi i}\,
2^{z_{3}-1}\,x^{-z_{1}-z_{2}-z_{3}}
\nn\\ &&\quad
\times\Gamma\left(\begin{array}{c}
-z_{1},\, 1+z_{1},\,3-z_{2},\,-2+z_{2},\,5-z_{1}-z_{2}-z_{3},\,-z_{3},\,2+z_{3},\,z_{1}+z_{2}+z_{3}\\
4,\,4-z_{2} \\
\end{array}\right),\label{3dmbint}
\end{eqnarray}
where $q_1=q_2=q_3=-1/4$.

Similarly to the analytic expression of \eqn{IIexplicit}, also in this case we have many terms that are singular 
in $x=1$ even though the full expression is well defined. Moreover in cases like this where we have a semi-analytic expression
we find that the analytic part and the remaining part expressed in terms of a three-fold MB integral are 
separately well defined in $x=1$. In particular for the case of the integral $\cII{r}(x,\ep;1,3;2,-1)$ in \eqn{secondexample}
we obtain the following limit:
\begin{eqnarray}
\lim_{x\to 1}\cII{r}(x,\ep;1,3;2,-1)&=&-\frac{1}{6}\,\frac{1}{\ep^3}+
\left(-\frac{607}{60}+\frac{40}{3}\,\log(2)\right)\,\frac{1}{\ep^2}+\bigg(\frac{77349}{14400}+\frac{509}{24}\,\zeta_2\nonumber\\
&&-\frac{3571}{45}\,\log(2)+\frac{40}{3}\,\log^2(2)+\textrm{MBint}[1]\bigg)\,\frac{1}{\ep}+\O(\ep^0),
\end{eqnarray}  
where $\textrm{MBint}[1]$ is given by
\begin{equation}
\textrm{MBint}[1]=0.329808.
\end{equation} 
This number is the result of the MB integral in \eqn{3dmbint}
with the choice $x=1$ obtained using the \texttt{MATHEMATICA}
package \texttt{MB.m} \cite{Czakon:2005rk}.
Finally we note that this example is representative for a small subset of the collinear nested integrals which have
these features. They are $\cII{i}(x,\ep;1,3;k,l)$ and $\cII{r}(x,\ep;1,3;k,l)$ with $k=-1,1,2$ 
and $l=-1$ and $\cIJ{}(x,\ep;1,3,1,3;k)$ with $k=-1$. 
The results for the pole structure of all the remaining cases of nested collinear
integrals are fully analytic.

\FIGURE{
\hspace*{20mm}
\includegraphics[scale=0.72]{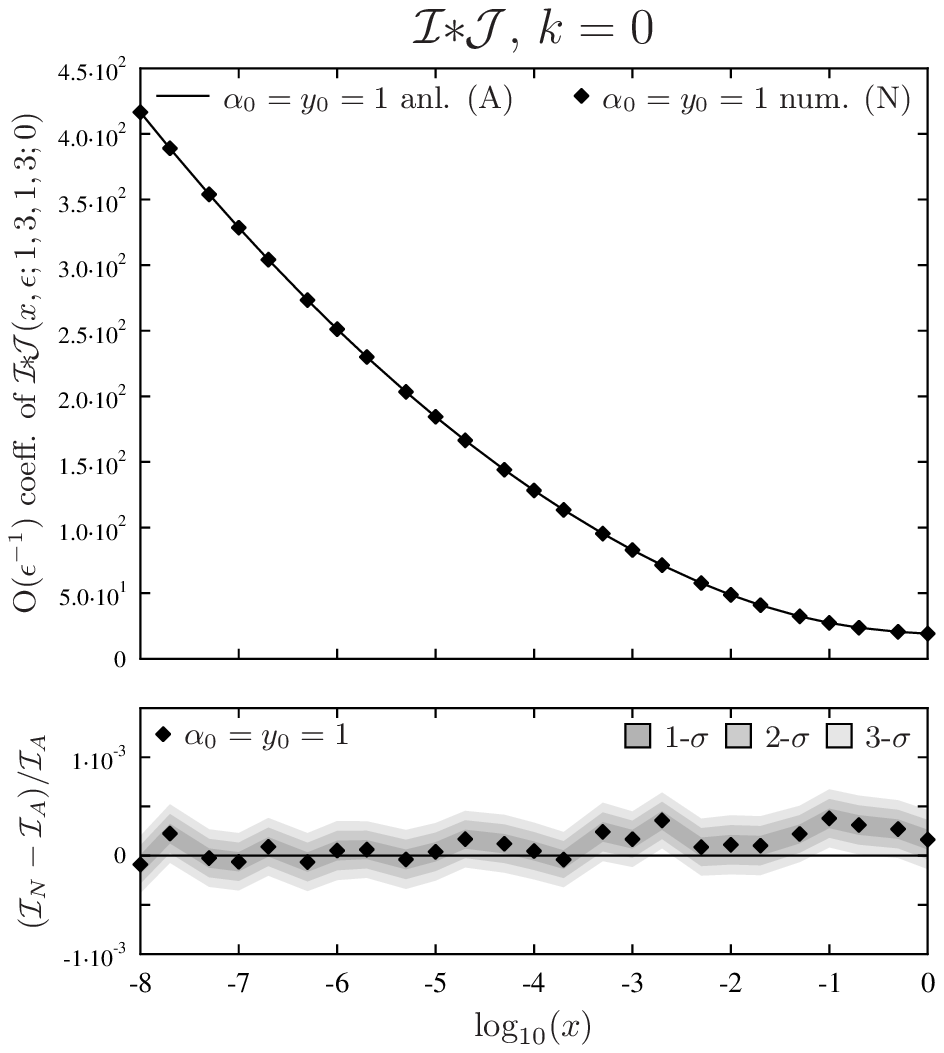}
\hspace*{20mm}
\caption{\label{fig:IJfigs}
Representative results for the ${\cIJ{}}$-type integrals. 
The plots show the coefficient of the $\Oe{-1}$ term for $k=0$ in 
$\cIJ(x,\ep;1,3,1,3;0)$ with $d_0=d_0'=3$ and $\alpha_0=y_0=1$.}
}

In \tab{tab:IInumbers} we list numerical values for the non-trivial coefficients 
of the $\ep$-poles (i.e. the
$\O(\ep^{-2})$ and $\O(\ep^{-1})$ coefficients) of the nested collinear integrals $\cII{r}(x,\ep;1,3;-1,2)$
 and $\cII{r}(x,\ep;1,3;2,-1)$. These numbers have been obtained using the fully
 analytic expression in \eqn{IIexplicit} and the semi-analytic one in
 \eqn{secondexample}. Numbers for the  $\O(\ep^{0})$ coefficient for the same
 representative integrals are listed in \tab{tab:IInumbers2}. In this case they have been
 entirely obtained evaluating their MB representations.

\TABLE{
\begin{tabular}{|l||c|c|c|c|}
\hline \hline
$\log_{10}(x)$ & \multicolumn{2}{|c|}
{\textbf{$\cII{r}(x,\ep;1,3;-1,2)$}} & \multicolumn{2}{|c|}
{\textbf{$\cII{r}(x,\ep;1,3;2,-1)$}} \\
\hline \hline
\,\,\,& $\O(\ep^{-2})$\, an. & $\O(\ep^{-1})$ \, an. & $\O(\ep^{-2})$ \, an. & $\O(\ep^{-1})$\, semi-an.\\
\hline \hline
-10. & -7.89751 & -374.957 & -15.9061 & -759.736 \\
\hline
-9.66667 & -7.64166 & -351.104 & -15.3944 & -711.688\\
\hline
-9.33333 & -7.38582 & -328.036 & -14.8828 & -665.211\\
\hline
-9. & -7.12998 & -305.753 & -14.3711 & -620.305\\
\hline
-8.66667 & -6.87413 & -284.256 & -13.8594 & -576.969\\
\hline
-8.33333 & -6.61829 & -263.544 & -13.3477 & -535.205\\
\hline
-8. & -6.36245 & -243.618 & -12.836 & -495.012\\
\hline
-7.66667 & -6.10661 & -224.477 & -12.3243 & -456.389\\
\hline
-7.33333 & -5.85076 & -206.122 & -11.8126 & -419.337\\
\hline
-7. & -5.59492 & -188.552 & -11.301 & -383.857\\
\hline
-6.66667 & -5.33908 & -171.768 & -10.7893 & -349.947\\
\hline
-6.33333 & -5.08324 & -155.769 & -10.2776 & -317.608\\
\hline
-6. & -4.82739 & -140.556 & -9.7659 & -286.841\\
\hline
-5.66667 & -4.57155 & -126.128 & -9.25423 & -257.644\\
\hline
-5.33333 & -4.31571 & -112.485 & -8.74256 & -230.019\\
\hline
-5. & -4.05986 & -99.628 & -8.2309 & -203.965\\
\hline
-4.66667 & -3.80402 & -87.556 & -7.71928 & -179.482\\
\hline
-4.33333 & -3.54818 & -76.269 & -7.20772 & -156.573\\
\hline
-4. & -3.29234 & -65.7665 & -6.69628 & -135.237\\
\hline
-3.66667 & -3.03649 & -56.0477 & -6.18508 & -115.476\\
\hline
-3.33333 & -2.78065 & -47.1111 & -5.6743 & -97.2944\\
\hline
-3. & -2.52481 & -38.9536 & -5.16432 & -80.6957\\
\hline
-2.66667 & -2.26896 & -31.5702 & -4.65576 & -65.6862\\
\hline
-2.33333 & -2.01312 & -24.9522 & -4.14969 & -52.2723\\
\hline
-2. & -1.75728 & -19.0853 & -3.64776 & -40.4585\\
\hline
-1.66667 & -1.50144 & -13.9478 & -3.15236 & -30.2408\\
\hline
-1.33333 & -1.24559 & -9.50936 & -2.66658 & -21.5965\\
\hline
-1. & -0.989751 & -5.73082 & -2.19382 & -14.4712\\
\hline
-0.66667 & -0.733908 & -2.5675 & -1.73699 & -8.76877\\
\hline
-0.33333 & -0.478065 & 0.0265877 & -1.2975 & -4.35339\\
\hline
0. & -0.222222 & 2.09462 & -0.874704 & -1.06702\\
\hline \hline
\end{tabular}
\caption{ Numerical values for the $\O(\ep^{-2})$ and $\O(\ep^{-1})$ coefficients of $\cII{r}(x,\ep;1,3;-1,2)$\,
(second and third column)\, and $\cII{r}(x,\ep;1,3;2,-1)$ \,(last two columns)\, for various 
values of $\log_{10}(x)$\, (first column). These numbers have been obtained evaluating the fully analytic expression in 
\eqn{IIexplicit} and the semi-analytic one in \eqn{secondexample} \label{tab:IInumbers}
}}

\TABLE{
\begin{tabular}{|l||c|c|}
\hline \hline
$\log_{10}(x)$ & 
\textbf{$\cII{r}(x,\ep;1,3;-1,2)$} & 
\textbf{$\cII{r}(x,\ep;1,3;2,-1)$} \\
\hline \hline
-5. & -1643.45 &  -3380.25\\
\hline
-4.66667 & -1354.81 & -2792.08\\
\hline
-4.33333 & -1104.32 & -2276.69\\
\hline
-4. & -886.741 & -1829.25\\
\hline
-3.66667 & -699.713 & -1444.98\\
\hline
-3.33333 & -541.331 & -1119.05\\
\hline
-3. &  -409.041 & -846.661\\
\hline
-2.66667 & -300.305 & -622.985\\
\hline
-2.33333 & -212.59 & -443.178\\
\hline
-2. & -143.341 & -302.315\\
\hline
-1.66667 & -89.9699 & -195.384\\
\hline
-1.33333 & -49.9194 & -117.263\\
\hline
-1. & -20.7583 & -62.7773\\
\hline
-0.66667 & -0.267788 & -26.8566\\
\hline
-0.33333 & 13.4889 & -4.81253\\
\hline
0. & 22.1523 & 7.37736\\
\hline \hline
\end{tabular}
\caption{\label{tab:IInumbers2} Numerical values for the $\O(\ep^{0})$ coefficient of $\cII{r}(x,\ep;1,3;-1,2)$\,
(second column)\, and $\cII{r}(x,\ep;1,3;2,-1)$ \,(last column)\, for various 
values of $\log_{10}(x)$\, (first column). These numbers have been obtained 
evaluating their MB representation. 
}
}

Finally in \fig{fig:IJfigs} we plot as a further example the fully analytic result for the first order
$\ep$-pole for $\cIJ(x,\ep;1,3;0)$ together with the numbers obtained numerically using
standard residuum subtraction and Monte Carlo numerical integration. As for all other
cases the agreement is excellent and the coefficient is given by a very smooth function of
$x$.


\section{Nested soft-type $\cJJ{}$ integrals}
\label{sec:JJints}

In this Section we discuss the analytic computation of the integrals defined in
\eqnss{eq:JJikint}{eq:JJkrint}. For them we were able to compute a fully analytic
result for the coefficient of the Laurent expansion up to $\O(\ep^{-2})$. 
The $\O(\ep^{-1})$ coefficient is computed semi-analytically similarly 
to the nested collinear integral  $\cII{r}(x,\ep;1,3 ;2,-1)$ discussed in  \sect{sec:IIints}. 
As a representative example
we show the structure of the fully analytic part of the result for the nested soft 
integrals $\cJJ{}$. For example choosing $d_0'=3$ we have:
\begin{eqnarray}
\cJJ{ik}(Y;\ep;1,3)&=&\frac{1}{\ep^{4}}+\left(\frac{22}{3}-2\,\log(Y)\right)\,\frac{1}{\ep^{3}}+
\cH(Y)\,\frac{1}{\ep^{2}}+\O(\ep^{-1}) 
\, , 
\label{eq:JJik}
\\
\cJJ{ir}(Y;\ep;1,3)&=&\frac{1}{2}\,\frac{1}{\ep^{4}}+\left(\frac{11}{3}-\log(Y)\right)\,\frac{1}{\ep^{3}}+
\bigg(\frac{533}{36}-\frac{22}{3}\,\log(Y)+\log^{2}(Y)
\label{eq:JJir}
\nn \\
&&
+\frac{3}{2}\,\textrm{Li}_{2}(1-Y)\bigg)\,\frac{1}{\ep^{2}} + \O(\ep^{-1})
\end{eqnarray}
and finally
\begin{equation}
\cJJ{kr}(Y;\ep;1,3)=\frac{1}{\ep^{4}}+\left(\frac{22}{3}-2\,\log(Y)\right)\,\frac{1}{\ep^{3}}+
\left(\cH(Y)-\zeta_2+\frac{1}{2}\,\textrm{Li}_{2}(1-Y)\right)\,\frac{1}{\ep^{2}}+\O(\ep^{-1})
\, .
\label{eq:JJkr}
\end{equation}
The function $\cH(Y)$ which appears in \eqns{eq:JJik}{eq:JJkr} is given by
\begin{equation}
\cH(Y)=\frac{497}{18}-2\,\zeta_2+\frac{6-8Y}{3\,(1-Y)^2}+\frac{33Y^3-117Y^2+126Y-44}{3\,(1-Y)^3}\,\log(Y)+2\,\log^{2}(Y)
+4\,\textrm{Li}_{2}(1-Y).\label{eq:refexpr}
\end{equation}
Also for this function even if some terms are singular at $Y=1$, we still have that the limit 
is well defined. Indeed we find
\begin{equation}
\lim_{Y\to 1}\cH(Y)=\frac{97}{3}-2\,\zeta_2.
\end{equation}

For these three soft-type integrals the $\O(\ep^{-2})$ coefficient has been plotted in
\fig{fig:JJfigs} using its fully analytic expression \eqns{eq:JJik}{eq:refexpr} and its numerical 
evaluation obtained using residuum subtraction and Monte Carlo integration. The agreement is
excellent and the analytic result confirms that also the coefficients of the Laurent
expansion for the $\cJJ{}$ integrals are smooth functions of $Y$.

\FIGURE{
\includegraphics[scale=0.72]{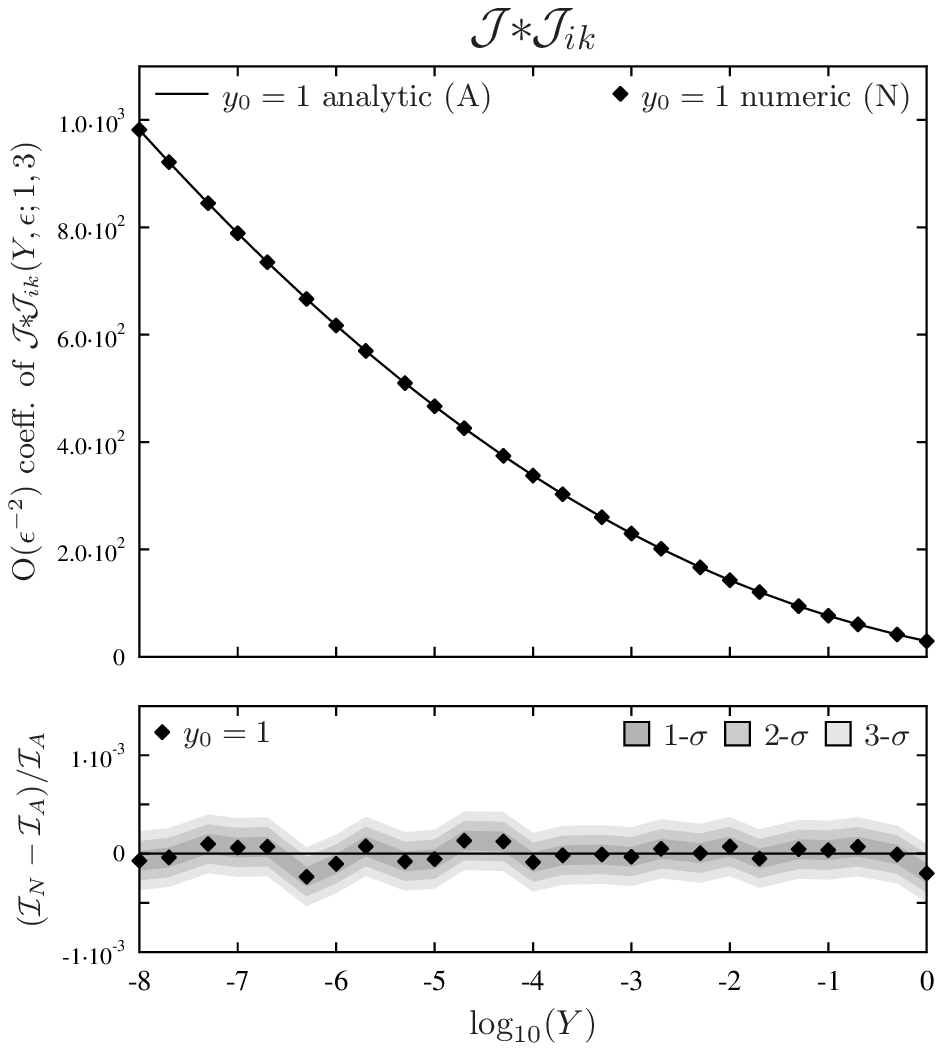}
\includegraphics[scale=0.72]{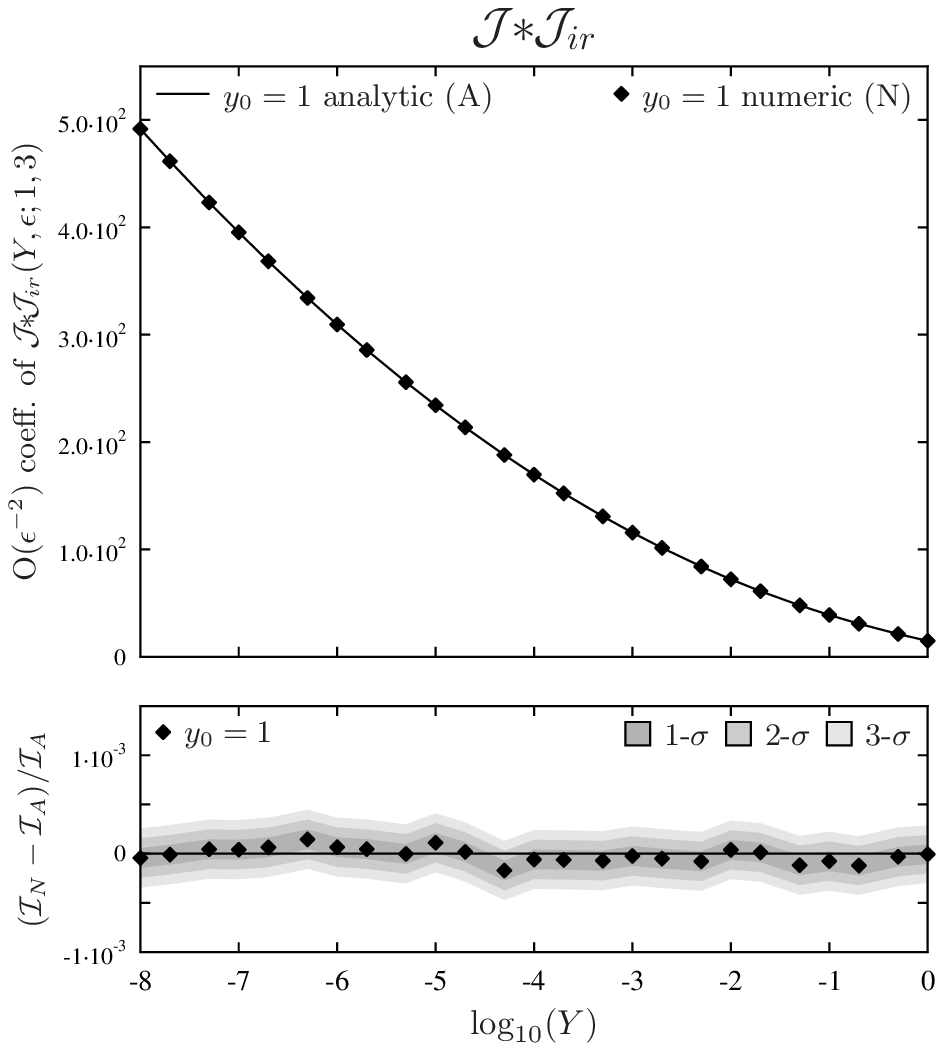}
\includegraphics[scale=0.72]{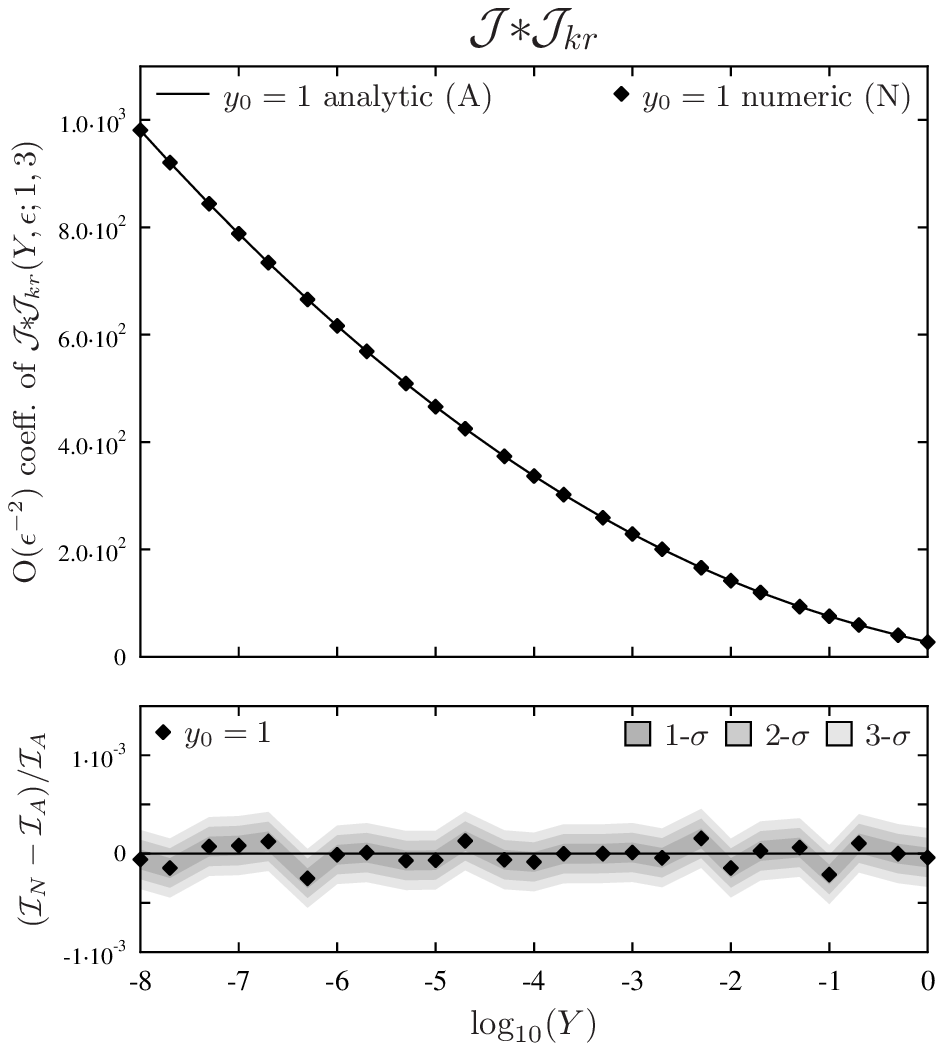}
\caption{\label{fig:JJfigs}
Representative results for the ${\cJJ{}}$-type integrals. 
The plots show the coefficient of the $\Oe{-2}$ term in 
$\cJJ{ik}(Y,\ep;1,3)$ (left), $\cJJ{ir}(Y,\ep;1,3)$ 
(right) and $\cJJ{kr}(Y,\ep;1,3)$ (bottom) with $d_0'=3$ and $y_0=1$.}
}

The numbers in \tab{tab:JJnumbers} have been obtained evaluating the nested soft integral
$\cJJ{ik}(Y;\ep;1,3)$ using the fully analytic expression in 
\eqn{eq:JJik} for the $\O(\ep^{-3})$ and $\O(\ep^{-2})$ coefficients. For the $\O(\ep^{-1})$
coefficient a semi-analytic expression in terms of a MB integral has been used and
finally the representation only in terms of MB integrals has been evaluated 
for the $\O(\ep^{0})$ coefficient.

\TABLE{
\begin{tabular}{|l||c|c|c|c|}
\hline \hline
$\log_{10}(Y)$ & \multicolumn{4}{|c|}
{\textbf{$\cJJ{ik}(Y;\ep;1,3)$}} \\
\hline \hline
\,\,\,& $\O(\ep^{-3})$\, an. & $\O(\ep^{-2})$\, an. & $\O(\ep^{-1})$ semi-an. & $\O(\ep^{0})$ MB\\
\hline \hline
-10. & 53.385 & 1430.99 & 25680. &  347094.\\
\hline
-9.66667 & 51.85 & 1350.22 & 23545.6 & 309328.\\
\hline
-9.33333 & 50.3149 & 1271.81 & 21533.4 & 274744.\\
\hline
-9. & 48.7799 & 1195.75 & 19639.8 & 243157.\\
\hline
-8.66667 & 47.2448 & 1122.05 & 17861.1 & 214389.\\
\hline
-8.33333 & 45.7098 & 1050.7 & 16193.8 & 188265.\\
\hline
-8. & 44.1747 & 981.714 & 14634.2 & 164617.\\
\hline
-7.66667 & 42.6396 & 915.081 & 13178.6 & 143283.\\
\hline
-7.33333 & 41.1046 & 850.805 & 11823.6 & 124106.\\
\hline
-7. & 39.5695 & 788.886 & 10565.3 & 106934.\\
\hline
-6.66667 & 38.0345 & 729.322 & 9400.38 & 91621.\\
\hline
-6.33333 & 36.4994 & 672.116 & 8325.04 & 78027.5\\
\hline
-6. & 34.9644 & 617.265 & 7335.7 & 66018.2\\
\hline
-5.66667 & 33.4293 & 564.771 & 6428.75 & 55463.9\\
\hline
-5.33333 & 31.8942 & 514.633 & 5600.58 & 46240.8\\
\hline
-5. & 30.3592 & 466.852 & 4847.56 & 38230.9\\
\hline
-4.66667 & 28.8241 & 421.427 & 4166.08 & 31321.6\\
\hline
-4.33333 & 27.2891 & 378.358 & 3552.51 & 25405.6\\
\hline
-4. & 25.754 & 337.645 & 3003.24 & 20381.7\\
\hline
-3.66667 & 24.219 & 299.287 & 2514.63 & 16153.6\\
\hline
-3.33333 & 22.6839 & 263.283 & 2083.06 & 12631.\\
\hline
-3. & 21.1488 & 229.632 & 1704.89 & 9728.88\\
\hline
-2.66667 & 19.6138 & 198.33 & 1376.45 & 7367.63\\
\hline
-2.33333 & 18.0787 & 169.37 & 1094.05 & 5473.13\\
\hline
-2. & 16.5437 & 142.739 & 853.961 & 3976.65\\
\hline
-1.66667 & 15.0086 & 118.417 & 652.369 & 2814.76\\
\hline
-1.33333 & 13.4736 & 96.3641 & 485.392 & 1929.49\\
\hline
-1. & 11.9385 & 76.5204 & 349.046 & 1268.34\\
\hline
-0.66667 & 10.4034 & 58.7892 & 239.262 & 784.581\\
\hline
-0.33333 & 8.86839 & 43.0286 & 151.932 & 437.509\\
\hline
0. & 7.33333 & 29.0435 & 82.998 & 192.684\\
\hline \hline 
\end{tabular}
\caption{Numerical values for the $\O(\ep^{-3})$, $\O(\ep^{-2})$, $\O(\ep^{-1})$ and $\O(\ep^{0})$ 
coefficients of $\cJJ{ik}(x;\ep;1,3)$ for various 
values of $\log_{10}(Y)$. The numbers  have been obtained from
\eqn{eq:JJik}, the semi-analytic one for the $\O(\ep^{-1})$ coefficient and MB integrals for the $\O(\ep^{0})$ coefficient.
  \label{tab:JJnumbers}
}}


\section{Nested soft-collinear $\cKJ{}$ integral}
\label{sec:KIint}

In this last Section we discuss the pole structure of the integral defined in \eqn{eq:KIint}. 
In this case the result for the
Laurent expansion is very simple because the integral has no dependence on the kinematics.
The coefficients of the poles in $\cKJ{}(\ep, 1,3)$ with $d_0'=3$ and $y_0=1$ read:
\begin{equation}
\cKJ{}(\ep, 1,3)=-\frac{1}{2}\frac{1}{\ep^{4}}-\frac{11}{3}\frac{1}{\ep^{3}}-\frac{557}{36}\frac{1}{\ep^{2}}
+\left(-\frac{10825}{216}+\frac{5}{3}\,\zeta_{2}-3\,\zeta_{3}\right)\frac{1}{\ep}+ \O(\ep^0).
\end{equation}
This completes our discussion of the analytic computation of the fundamental integrals that contribute 
to the singly-unresolved counterterms.


\section{Conclusions}
\label{sec:conclusion}

In this work we have completed the evaluation of all integrals needed 
for the computation of the integrated real-virtual counterterms 
of the subtraction scheme for NNLO jet cross sections 
proposed in \Refs{Somogyi:2006cz,Somogyi:2006db,Somogyi:2006da}. 
We have discussed representative examples for all types of soft and collinear as well as nested integrals 
in \sectss{sec:Iints}{sec:KIint} (the complete results are contained in a \texttt{MATHEMATICA} file).
These integrals (i.e. their Laurent expansions in $\ep$ to sufficient depth) 
have to be computed once and for all and their knowledge is necessary in order to make the subtraction scheme an effective tool.  
We have achieved this task by deriving MB representations for all integrals under consideration and, in a subsequent step, 
we have performed analytically the summation of the nested sums over the series of residues.
In some cases, this second step of summing the series has not been achieved and we have resorted to a numerical
evaluation of the MB integrals in the complex plane.
As a further check, all MB representations for both the numerical and, if available, the analytic results have
also been compared against an independent evaluation of the integrals using standard residuum subtraction together with 
the Monte Carlo integration in \Ref{Somogyi:2008fc}. 
We have shown, that all integrals contributing to the real-virtual counterterms are smooth functions. 
For practical applications, this means that all integrals (in particular the
finite in $\ep$ contributions) can be used in terms of interpolating tables, which are computed once and for all.

Files of our results can be obtained from the preprint server {\tt http://arXiv.org} by downloading the source. 
They are also available at~\cite{Zeuthen-CAS:2009} or from the authors upon request.

\bigskip


\acknowledgments{
We acknowledge useful discussions with J.~Bl\"umlein, T.~Riemann and V.~Yundin .
This work is supported in part by the
Deutsche Forschungsgemeinschaft in SFB/TR 9, 
the Helmholtz Gemeinschaft under contract VH-NG-105, 
the Hungarian Scientific Research Fund grand OTKA K-60432 and by 
the Swiss National Science Foundation (SNF) under contract 200020-117602. 
}



\providecommand{\href}[2]{#2}\begingroup\raggedright\endgroup

\end{document}